\documentclass[12pt]{article}
\usepackage[margin=2cm]{geometry}
\usepackage{amsmath}
\usepackage{xcolor}
\usepackage{graphicx,mathrsfs}

\begin{document}
\title{Exciton absorption spectra in
narrow armchair graphene nanoribbons in electric fields}
\author{ B.~S.~Monozon\\
Physics Department, Marine Technical University, 3 Lotsmanskaya Str.,\\
190121 St.Petersburg, Russia,\\
\\
P.~Schmelcher\\
Zentrum f\"ur Optische Quantentechnologien, Universit\"{a}t Hamburg, \\ Luruper Chaussee 149, 22761 Hamburg, Germany\\
The Hamburg Centre for Ultrafast Imaging, Universit\"{a}t Hamburg,~\\ Luruper Chaussee 149, 22761 Hamburg, Germany}
\date{\today}
\maketitle

\begin{abstract}
We present an analytical investigation
of the exciton optical absorption
in a narrow armchair graphene nanoribbon (AGNR)
in the presence of a longitudinal external electric field
directed parallel to the ribbon axis.
The two-body 2D Dirac equation for the massless
electron and hole
subject to the ribbon confinement, Coulomb interaction
and electric field is employed.
The ribbon confinement is assumed to be much
stronger than the internal
exciton electric field
which in turn considerably exceeds the external
electric field. This justifies an
adiabatic approximation implying a slow longitudinal
and fast transverse electron-hole relative motion
governed by the exciton attraction and
accompanied by the external electric field and ribbon confinement effects, respectively.
In the single subband approximation of the
isolated size-quantized subbands
induced by the ribbon confinement
the exciton
electroabsorption coefficient is determined in
an explicit form. The pronounced
dependencies of the exciton peak positions,
widths and intensities on the ribbon width
and electric field strength are traced.
The electron-hole exciton attraction
modifies remarkably the Franz-Keldysh electroabsorption
in the frequency region both below and above the edges determined
by the size-quantized energy levels.
In the double-subband
approximation of the interacting ground and
first excited subbands the combined effect of the exciton
electro- and autoionization caused
by the electric field and intersubband coupling,
respectively, are explored.
Our analytical results are in agreement
with those obtained by numerical methods. Estimates of the expected experimental values
for the typically employed AGNR show that for a weak electric field
the exciton quasi-discrete states remain sufficiently stable
to be observed in optical experiments,
while relatively strong fields free the captured
carriers to further restore their contribution to the transport.
\end{abstract}

\section{Introduction}\label{S:intro}

Following the early work in Ref. \cite{ Wakab96}
especially
during the last decade a large number of
experimental and theoretical
studies have been performed on
the transport, electronic, and optical properties
of armchair graphene nanoribbons (AGNR)
(see \cite{ Wakab, Alf, Soavi, Jia} and references therein).
AGNR are spatially confined elongated strips of graphene monolayer.
This is in particular due to
the fact that in contrast to
gapless unbound 2D graphene monolayers,
AGNR
are favorable for the
formation of excitons
(bound electron-hole pairs), which in turn
strongly affect the 2D graphene electronic and optical properties.
In 2D graphene
the vanishing density of the electron and hole states
at the Dirac points prevents the formation of
excitons, while in the quasi-1D AGNR
possessing an open band gap
the bound exciton states can arise \cite{Castnet}.
The existence of these states provides several reasons
for the strong interest in AGNRs.
The first reason is that AGNR being the interconnects between
2D graphene monolayers must support their ultrahigh
carrier mobility in graphene-based nanoelectronic devices. At the same
time the exciton effect, i.e. electron-hole attraction transforms free electrons and holes
into neutral excitons and thereby suppresses
the charge transport properties of graphene monolayers.
Note that excitons in AGNR are bound much stronger
than their counterparts in quasi-1D semiconductor
structures, namely in the quantum wire (QWR) and in a bulk crystal subject
to a strong magnetic field (diamagnetic excitons (DE)).
The binding energies $E^{(\mbox{\tiny{b}})}$ of the excitons in the AGNR
of width $d \simeq 1~ \mbox{nm}$ reach the considerable
value of $E^{(\mbox{\tiny{b}})} \simeq 1 \,\mbox{eV}$ \cite{Yang}, while
for realistic magnetic fields and QWR \cite{Zhang, Mad}
$E^{(\mbox{\tiny{b}})} \simeq 10-20\, \mbox{meV}$.
Thus, the liberation
mechanism
for the AGNR charge carriers captured by the excitons
is a relevant question of immediate interest.

In addition, the exciton
effect changes drastically the optical properties of the
AGNR. In particular, this is the absorption in the vicinity of the
subband edges determined by the size-quantized electron-hole
energy levels. The square-root
divergency of the fundamental absorbtion profile
specific for the quasi-1D structures
is replaced by the finite value, associated with the peaks of the Rydberg series peaks broadening
below each edge.

The second reason is that quasi-1D
structures, namely DE \cite{Zhilmak}, QWR
\cite{Monschm09} and
AGNR \cite{Monschm16} are favorable for the formation of not only strictly
discrete but also metastable Fano resonances \cite{Fano} like exciton
states adjacent to the excited size-quantized energy levels.
The nature of the Fano resonances lies in the
inter-subband coupling between the discrete and low-lying continuous
Rydberg states. Narrow AGNR outperform the mentioned 1D semiconductor
structures in terms of the impact of the
binding energy of the exciton states, their resonant energy widths and
related parameters. The exciton binding energies
and widths
are proportional to the exciton
Rydberg constant $Ry$ \cite{Monschm16}.
For the semiconductor structures the Rydberg constant does not
depend on the confinement parameters, i.e.
on the radius \cite{Monschm09}
and magnetic field \cite{Hashow} for the
QWR and DE, respectively. For the AGNR with width $d$
we have $Ry\sim d^{-1}$ \cite{Nov}.
Thus, for narrow AGNRs dimensionality plays
a crucial role and allows to control the relevant
exciton properties.

External electric fields directed parallel to the graphene
axis strongly affect
the excitonic and related carrier mobility, the Fano resonances
and optical absorption spectrum.
The process of electric field ionization
can be used as a tool for the release of the carriers bound
within the strictly discrete and quasi-discrete exciton states
and thereby enhance the conductance of both the AGNR
and the nanoelectronic devices these ribbons
are incorporated into. As established earlier
for the bulk semiconductors, the electric fields
act significantly on both the fundamental
 (Franz-Keldysh effect \cite{Frkeld})
and exciton \cite{Knox} absorption. They shift and broaden
the discrete exciton peaks
and modify remarkably the exciton spectrum both below \cite{Merper}
and above \cite{Arios} the optical edge.

In addition, quasi-1D excitons in AGNR subject to electric fields
allow us to study the process of exciton double channel
ionization.
The excited metastable exciton Rydberg series decay
according to
two channels: the autoionization channel, open due to the intersubband
Fano coupling \cite{Zhilmak, Fano},
and the channel of electric field ionization,
caused by under barrier tunneling \cite{Silv}.

Note that the majority of the theoretical approaches to the problem
of the exciton in GNR are based on
numerical calculations implying a considerable computational
effort. The discrete part of the exciton absorption spectrum
of GNR has been calculated numerically \cite{Yang, Prez, Zhu, Denk}
using density functional theory, employing the local density approximation, and the
Bethe-Salpeter equation. Jia \emph{et al.} \cite{Jia}
and Lu \emph{et al.} \cite{Lu} used the tight-binding approximation,
while Alfonsi and Meneghetti \cite{Alf} employed a full many-body exact
diagonalization of a parametric Hubbard Hamiltonian in their
calculations of the exciton peak positions and intensities.
Gundra and Shukla \cite{Gund} studied computationally
the optical absorption
in the zigzag GNR in the presence of electric fields.
The Pariser-Parr-Pople-model Hamiltonian was employed, however the exciton effect was not taken into account. Only few works rely on analytical methods
\cite{Wakab, Monschm16, Nov, Mohamm}, while many
numerical studies of the electronic and optical properties
of the GNR have been performed
(see Ref. \cite{Gund} and references therein). Numerical results are very helpful also
for the detailed interpretation of a specific experiment.
At the same time analytical approaches are appropriate
to reveal the basic physics of the underlying phenomena  especially at the first stages of the investigation.
To our
knowledge, analytical studies of the exciton electroabsorption
in AGNR have not been performed in the literature yet.

In the present work, we develop an analytical approach
to the problem of the exciton states in the narrow AGNR
in the presence of an external electric fields
directed parallel to the ribbon axis. The Coulomb
electron-hole attraction is taken to be much weaker
than the influence of the ribbon confinement
and much stronger than the effect of the electric fields.
The two-body 2D Dirac equation for the massless
electron and hole in AGNR
subject to the Coulomb
and external electric
fields is solved in the adiabatic approximation.
This approximation implies that the transverse
electron-hole motion governed by the ribbon
confinement is much faster than the longitudinal
motion controlled by the Coulomb and external electric
fields. In the approximation of the
isolated size-quantized subbands
the optical absorption coefficient in the vicinity
of the quasi discrete exciton states
(peak positions, widths and intensities)
and within the continuous band
as a function of
the ribbon width and electric field strength are determined explicitly.
In the double-subband approximation the total widths
of the first excited Rydberg series exciton
peaks, associated with the electric field ionization
and intersubband Fano coupling
are investigated.
Numerical estimates of the expected experimental values
for realistic AGNR parameters and electric field strenghts are made.
The aims of this work are to study the effect of the
electric field and ribbon confinement on the
exciton absorption spectrum and
to elucidate the mechanism
of the
ionization process of the excitons yielding the increase
of the carriers mobility in the AGNR.
In addition, we intend to trigger further experimental
and theoretical studies.

The paper is organized as follows.
In Section 2 the general analytical equations are derived.
The exciton electroabsorption coefficient is determined
in the single- and double-subband approximations in Sections
3 and 4, respectively. A discussion of the obtained theoretical
results and estimates of the expected experimental values
is presented in Section 5. Section 6 contains our conclusions.

\section{General approach}\label{S:gen}

We consider the exciton optical absorption in
electrically biased AGNR with width $d$
and length $L$ placed on the $x-y$ plane and bounded by
straight lines $x=\pm d/2.$
The uniform electric field
$\vec{F}$ as well as the polarization of
the involved light wave
are chosen to be parallel to the ribbon $y$-axis.
The general approach to the problem of the
exciton absorption is based on Refs.
\cite{Sas} and \cite{Ell} devoted to the
interband optical absorption in AGNR and
exciton absorption in semiconductors, respectively.
Since the mathematical details of this approach have been presented
in Ref. \cite{Monschm16}
only an outline of our calculations will be provided below.
The equation for the exciton absorption coefficient has the form

\begin{equation}\label{E:abs}
\alpha = \sum_N \alpha^{(N)};\quad
\alpha^{(N)} = \frac{1}{n_b \varepsilon_0 c}\sigma_{yy}^{(N)},
\end{equation}
where $n_b$ is the refractive index of the ribbon substrate,
$c$ is the speed of light, and
$\sigma_{yy}^{(N)}$ is the component of the dynamical conductivity,

\begin{equation}\label{E:cond}
\sigma_{yy}^{(N)}=  \frac{\pi p^2 e^2}{\hbar S \Delta_{N}}
\sum_{n,s}\big |\sigma^{(N)}_{xn(s)}\big |^2 \delta \left(\hbar \omega - E_{Nn(s)}\right)
\delta_{\vec{q}_{ph} \vec{K}}
\end{equation}
determined by the matrix element

\begin{equation}\label{E:matr}
\sigma^{(N)}_{xn(s)} = \left <\vec{\Psi}^{(0)}(\vec{\rho}_e ,\vec{\rho}_h)
|\hat{\sigma}_{xh}\bigotimes \hat{I}_e + \hat{I}_h \bigotimes \hat{\sigma}_{xe}|
\vec{\Psi}^{(\mbox{\tiny{x}})}_{Nn(s)}(\vec{\rho}_e ,\vec{\rho}_h)   \right >
\end{equation}
of the Pauli matrix $\hat{\sigma}_{x}$ calculated between the ground state
$\vec{\Psi}^{(0)}$ and exciton wave functions $\vec{\Psi}^{\mbox{\tiny (x)}}_{Nn(s)}$
of the bound $(n)$ and continuous $(s)$ states of the exciton, formed
by an electron and hole related to size-quantized energy subbands
with the common index $N$.
The exciton states consisting
of the electron and hole associated with the different
$N_e \neq N_h$ subbands are optically inactive and can be excluded from
expansion (\ref{E:Psi})
(see justification in \cite{Monschm05} and references therein).
 As usual, the symbol $\bigotimes$ denotes the tensor
product of the Pauli $\hat{\sigma}_{x}$ and unit $\hat{I}$ matrices. In eq. (\ref{E:cond})
$p=\hbar v_F~(v_F = 10^6\,\mbox{m/s})$ is the graphene energy parameter,
$S=Ld$ is the area of the ribbon, $\Delta_N = 2\varepsilon_N$ is the
effective energy gap between the electron and hole subbands,
branching from the size-quantized levels $\pm \varepsilon_N$
in the conduction and valence bands,
respectively. The $\delta$-functions in eq. (\ref{E:cond})
reflect the conservation laws in the system
formed by the absorbed photon with the energy $\hbar \omega$
and momentum $\hbar\vec{q}_{\mbox{\tiny {ph}}}$ plus the emersed exciton of the energy
$E_{Nn(s)}$ and total momentum $\hbar\vec{K}$.

Following Elliot's approach justified in detail in Ref.
\cite{Ell} the wave function $\vec{\Psi}^{(0)}$
of the ground state of the electron-hole pair
in a semiconductor-like AGNR is chosen in the form

\begin{equation}\label{E:grfun}
\vec{\Psi}^{(0)}(\vec{\rho}_e ,\vec{\rho}_h)=\delta (y)\delta (x_e -x_h)
\left[\vec{\Phi}_A^{(0)}\bigotimes\vec{\Phi}_A^{(0)} +
 \vec{\Phi}_B^{(0)}\bigotimes\vec{\Phi}_B^{(0)}\right],
\end{equation}
where $y=y_e -y_h$ is the relative $y$-coordinate and

$$
\vec{\Phi}_A^{(0)} =\frac{1}{\sqrt{2}}
\begin{array}{c}
\begin{Bmatrix}
-1 \\
0 \\
1\\
0
\end{Bmatrix}
\end{array};
\qquad
\vec{\Phi}_B^{(0)} =\frac{1}{\sqrt{2}}
\begin{array}{c}
\begin{Bmatrix}
0 \\
1 \\
0\\
-1
\end{Bmatrix}
\end{array}.
$$
The exciton wave function $\vec{\Psi}^{\mbox{\tiny (x)}}$
obeys the equation

\begin{equation}\label{E:basic}
\hat{{\rm H}}_{\mbox{\tiny x}}\vec{\Psi}^{\mbox{\tiny (x)}}(\vec{\rho_e }, \vec{\rho_h })=
E\vec{\Psi}^{\mbox{\tiny(x)}}(\vec{\rho_e }, \vec{\rho_h }).
\end{equation}
In this equation

\begin{equation}\label{E:exham}
\hat{{\rm H}}_{\mbox{\tiny x}} = \hat{{\rm H}}_h (\hat{\vec{k}}_h)
\bigotimes\hat{I}_e + \hat{I}_h \bigotimes \hat{{\rm H}}_e (\hat{\vec{k}}_e)
+\hat{I}_h\bigotimes \hat{I}_e
\left[V(\vec{\rho_e} -\vec{\rho_h })- eF(y_e -y_h)\right]
\end{equation}
is the traditional exciton Hamiltonian \cite{Mohen} formed by the electron
and hole Hamiltonians $\hat{{\rm H}}_j (\hat{\vec{k}}_j),~j=e,h$ corresponding
to the nonequivalent Dirac points\\
$\vec{K}^{(+,-)}= \pm K, 0;~(K = 4\pi / 3a_0 ,~a_0 = 2.46\, \mbox{{\AA}}$
is the graphene lattice constant) \cite{BreyFert} and

\begin{equation}\label{E:coul}
V(\vec{\rho}_e,\vec{\rho}_h)=
-\frac{e^2}{4\pi\varepsilon_0 \epsilon_{\mbox{\tiny eff}}\sqrt{(x_e-x_h)^2 +(y_e-y_h)^2 }}.
\end{equation}
is the 2D Coulomb potential of the electron-hole attraction.
Here $\epsilon_{\mbox{\tiny eff}}=\frac{1}{2}(1+\epsilon + \pi q_0)$ is the effective
dielectric constant determined by the static dielectric constant $\epsilon$ of the
substrate and by the parameter $q_0 = e^2/4\pi \varepsilon_0 p \simeq 2.2 $ \cite{Nov, Hwang}.

Further we choose the exciton wave function $\vec{\Psi}^{\mbox{\tiny (x)}}$ in the form

\begin{equation}\label{E:Psi}
\vec{\Psi}^{\mbox{\tiny (x)}}(\vec{\rho_e }, \vec{\rho_h })=
\frac{1}{\sqrt{2}}\sum_N
\sum_{\alpha =A,B} u_{N\alpha}(y_h)\vec{\Phi}_{N\alpha}(x_h)\bigotimes
\sum_{\beta =A,B}u_{N\beta}(y_e)\vec{\Phi}_{N\beta}(x_e),
\end{equation}
where $\vec{\Phi}_{N A(B)}(x_{e(h)}) ~\mbox{and}~ u_{N A(B)}(y_{e(h)})$
are the wave functions describing the electron (hole)
transverse $x-$ and longitudinal $y-$ states, governed by the
ribbon confinement and exciton $V(\vec{\rho}_e -\vec{\rho}_h)$
(\ref{E:coul}) and electric field $-eF(y_e - y_h)$ potentials, respectively,
in the $A (B)$ graphene sublattices.
In equation (\ref{E:Psi}) the sublattice wave functions
$\vec{\Phi}_{NA,(B)}$ are as follows

$$
\vec{\Phi}_{NA}(x_j) =
\begin{array}{c}
\begin{Bmatrix}
-\varphi_N(x_j) \\
0 \\
\varphi_N^{\ast}(x_j)\\
0
\end{Bmatrix}
\end{array};
\qquad
\vec{\Phi}_{NB}(x_j) =
\begin{array}{c}
\begin{Bmatrix}
0 \\
\varphi_N(x_j) \\
0\\
-\varphi_N^{\ast}(x_j)
\end{Bmatrix}
\end{array},\quad j=e,h,
$$
where the explicit form of the functions $\varphi_N(x_j)$
is presented in Ref. \cite{Monschm12}.
The electron (hole) energies
corresponding to the size-quantized transverse $N$ states
are equal to $+(-)\varepsilon_N$ with

\begin{equation}\label{E:ener}
\varepsilon_N=|N-\tilde{\sigma}|\frac{\pi p}{d};~N=0,\pm1,\pm2,\ldots~;
\quad\tilde{\sigma}=\frac{Kd}{\pi}-\left[ \frac{Kd}{\pi}\right].
\end{equation}
Below, to be specific, we will consider AGNR of the family $\tilde{\sigma} = 1/3$, providing
a semiconductor-like gap structure.

Substituting the expansion of the wave function
$\vec{\Psi}^{\mbox{\tiny (x)}}(\vec{\rho_e }, \vec{\rho_h })$
 over the ortho-normalized basis set
$\vec{\Phi}_{N\alpha}(x_h)\bigotimes\vec{\Phi}_{N\beta}(x_e)$
(\ref{E:Psi}) into eq. (\ref{E:basic}) and in view of the equations
$$
\hat{H}_j (\hat\vec{k_j})\vec{\Phi}_{NA(B)} (x_j)=\varepsilon_N \vec{\Phi}_{NB(A)}(x_j)
$$
we arrive after routine manipulations to the set of the equations for the expansion
coefficients
$u_{N\alpha}(y_h) u_{N\beta}(y_e)$
(see eq. (14) in Ref. \cite{Monschm16})
written in terms of the centre of mass
$Y=\frac{1}{2}(y_e + y_h)$ and relative $y = y_e - y_h$ coordinates

$$
u_{N\alpha}(y_h) u_{N\beta}(y_e)= \frac{{\rm e}^{{\rm i}QY}}{\sqrt{L}}\xi_{N\alpha\beta}(y)~;\,
\xi_{NAA}=\xi_{N1}~;\,  \xi_{NAB,(BA)}= \frac{1}{\sqrt{2}}(\xi_{N2}\pm \xi_{N3})~;\,
\xi_{NBB}=\xi_{N4};
$$
$\hbar Q$
is the longitudinal component of the exciton total momentum.

Further this set is solved in the adiabatic approximation.
This implies that the fast transverse $x$- and slow $y$- motions
affected by the ribbon confinement and exciton attraction plus weak electric field, respectively, are adiabatically separated.
The adiabaticity parameter
$q$ i.e. the Coulomb potential strength (\ref{E:coul})
scaled with the graphene energetic parameter $p$
and the imposed adiabatic condition are as follows

\begin{equation}\label{E:adiab}
q=\frac{e^2}{4\pi\varepsilon_0 \epsilon_{\mbox{\tiny eff}}p};\quad q << 1.
\end{equation}
Under this condition the set for the functions
$\xi_{Nj},~j=1,2,3,4.$ \cite{Monschm16} transforms into that
for the functions
$\xi_{N1} = \xi_{N4}=
\frac{1}{\sqrt{2}}\xi_{N2} \equiv \xi_{N},~ \xi_{N3} \ll \xi_{N1}$

\begin{equation}\label{E:main}
-\frac{\hbar^2}{2\mu_N}\xi_{N}^{''}(y) +
\left[V_{NN}(y) - eFy - W_N \right]\xi_{N}(y) +
 \sum_{N'\neq N}V_{N'N}(y)\xi_{N'}(y)=0.
\end{equation}
These equations describe the relative motion
of the 1D exciton with centre-of-mass
momentum $Q=0$, reduced mass
$\mu_N =\frac{\hbar^2\Delta_N }{4p^2}$
and energy $W_N = E - \Delta_N $ in the presence of
the external electric $-eFy$ and quasi-Coulomb $V_{N'N}(y)$
potentials, where

\begin{align}\label{E:1Dc}
V_{N'N}(y)=\frac{1}{d^2}\int_{-\frac{d}{2}}^{+\frac{d}{2}}dx_e
\int_{-\frac{d}{2}}^{+\frac{d}{2}}dx_h V(\vec{\rho})
\cos \left[(N-N')\pi \left(\frac{x_e}{d}- \frac{1}{2}\right)  \right]
\\
\nonumber
\times
\cos \left[(N-N')\pi \left(\frac{x_h}{d}- \frac{1}{2}\right)  \right]~;\,\vec{\rho} = \vec{\rho}_e - \vec{\rho}_h~,
\end{align}
determined by eq. (\ref{E:coul}) for the potential
$V(\vec {\rho})$ with

\begin{equation}\label{E:onoff}
V_{N'N}(y)=-\frac{\beta}{|y|}\left[ \delta_{N'N}+O \left(\frac{d^2}{y^2}  \right)
\delta_{|N'-N|(2s+1)} \right]~;\,s=0,1,2,\ldots ;~\mbox{at}~|y|\gg d.
\end{equation}
Other parameters
related to the $N$th subband are the exciton Bohr radius
$a_N = \frac{4\pi \varepsilon_0 \epsilon_{\mbox{\tiny eff}}\hbar^2}{\mu_N e^2}$,
exciton Rydberg constant
$Ry_N = \frac{\hbar^2}{2\mu_N a_N^2}(=\frac{\Delta_N q^2}{8})$
and dimensionless electric field $f_N = \frac{F}{F_N}$,
which is the external electric field $F$ scaled with the exciton
electric field $F_N = \frac{Ry_N}{ea_N}$.

Using eqs. (\ref{E:grfun}) and (\ref{E:Psi}) for the  ground
state $\vec{\Psi}^{(0)}$ and exciton
$\vec{\Psi}^{(\tiny x)}$ wave functions,
respectively, we determine the matrix element (\ref{E:matr})
of the dipole exciton optical transition in the form
$\sigma_{xn(s)}^{(N)}= - \sqrt{L}\xi_{N}(0) $. As expected,
for the noninteracting electron-hole pair for which
$\xi_{N}(y)= \frac{1}{\sqrt{L}}\exp^{{\rm i} sy}$
the matrix element of the fundamental optical transition
reads $|\sigma_{xn(s)}^{(N)}|=1$ \cite{Sas}. The contribution
$\alpha^{(N)}$ (see eqs. (\ref{E:abs}), (\ref{E:cond})) to the
coefficient of the exciton absorption $\alpha$
in the vicinity of the edge
$\Delta_N $ takes on the following appearence

\begin{equation}\label{E:absgen}
\alpha_N (\omega) = \alpha^{(0)}\frac{4\pi p^2}{n_b \Delta_N d}
\sum_{n(s)}\big| \xi_{N1}(0) \big|^2
\delta \left(\hbar \omega - E_{Nn(s)}\right),
\end{equation}
where $\alpha^{(0)} = e^2/4\varepsilon_0 \hbar c\simeq 2.3\cdot10^{-2}$
is the absorption of the suspended graphene.
As mentioned above the details of the calculations of eq.
(\ref{E:absgen}) including the explicit forms of the
Hamiltonians $\hat{H}_{e,h}(\hat{\vec{k}}_{e,h})$ in eq. (\ref{E:exham}),
sublattice wave functions $\vec{\Phi}_{NA,(B)}(x)$
in eq. (\ref{E:Psi})
and set of functions $\chi_{N1} - \chi_{N4}$
can be found in Ref. \cite{Monschm16}.

\section{Spectrum of the exciton electroabsorption:
Single-subband approximation}\label{S:spectr}

Here we employ the single-subband approximation ignoring the coupling
between the electron-hole subbands with the
different indices $N$. It follows from eq. (\ref{E:onoff})
that in the narrow ribbon
with small width $d$ the diagonal potentials $V_{NN}$ dominate the
off-diagonal ones in the set of equations (\ref{E:main}). This allows us to set $V_{N'N} = V_{N}\delta_{N'N}$,
omit the $\Sigma_{N'}$ from eqs. (\ref{E:main}) and arrive
at the equations for the functions $\chi_{N}(y)$
governed by the diagonal Coulomb contribution

\begin{eqnarray}\label{E:diag}
V_{N}(y)=\frac{2}{d}qp
\left[\ln\frac{\frac{|y|}{d}}{1 + \sqrt{1+\frac{y^2}{d^2}}}+
\sqrt{1+\frac{y^2}{d^2}} - \frac{|y|}{d}\right]=
\left\{
\begin{array}{cl}
\frac{qp}{d}\ln\frac{y^2}{d^2}~;\, &\frac{|y|}{d}\ll 1\\
-\frac{qp}{|y|}~;\, &\frac{|y|}{d}\gg 1
\end{array}
\right.
\end{eqnarray}
and the electric field potential $-eFy$.

These equations are solved by matching in the
intermediate regions the wave functions valid in the inner
$0\leq \mid y \mid \ll a_N$,
Coulomb $d\ll \mid y\mid \ll (\mid y_N \mid   a_N)^{\frac{1}{2}}$
and "electric" $\mid y \mid \gg (\mid y_N \mid   a_N)^{\frac{1}{2}}$
regions, determined by the exciton Bohr radius $a_N$ and turning point
$y_N = -\frac{W_N}{eF}$ for vanishing classical momentum
$\mathscr{P}(y) = \sqrt{2\mu_N \left(W_N +eFy\right)}$.
In the inner and Coulomb regions the exciton electric field $F_N$
considerably exceeds the external electric field $F$, while in the
"electric" region the exciton potentials $V_{N}(y)$ can be treated
as a small perturbation to the external field effects.
The exciton absorption coefficient is determined
for the photon energies $\hbar\omega =E$ below the absorption edge
$0>\hbar \omega - \Delta_N \simeq - \frac{Ry_N}{n^2},
~n\simeq 0,1,2,\ldots$  close to the energies of the
bound Rydberg states, for the frequencies
$\hbar \omega - \Delta_N < 0$ positioned far away from the $Nn$
exciton peaks and for the frequency region above the edge $\hbar \omega - \Delta_N > 0.$
The approach presented in this section closely resembles those
in the works \cite{Monschm09, Monschm14} dedicated to the exciton electroabsorption
in QWR and impurity states
in AGNR subject to electric fields, respectively. Here we focus only on the main features
leaving aside the details.

\subsection{Frequency region $\hbar \omega - \Delta_N < 0 $}

At this stage it is convenient to introduce the exciton
quantum number $\kappa_N$
and reciprocal length $\nu_N$ defined by
$W_N = - Ry_N/\kappa_N^2 $
and $\nu_N = 2(\kappa_N a_N)^{-1}$, respectively.
In the inner region
an iteration method is employed. Double integration
of eq. (\ref{E:main}) without the term involving
$\sum_{N'\neq N}$
with the trial function
and its derivative $\xi_N^{(0)}(y)=c_N, ~ \xi_N^{(0)'}(y)=0$
relevant to the optically active excitons
(see eq. (\ref{E:absgen}) ) generates the function

\begin{equation}\label{E:iter}
\xi_N (y) = c_N \left[1-
2 \frac{y}{a_N}\left(\ln \frac{2y}{d} - \frac{1}{2}   \right) \right].
\end{equation}
In the Coulomb region
the general solution to eq. (\ref{E:main})

\begin{equation}\label{E:coulomb}
\xi_N (y) = A_N W_{\kappa_N , \frac{1}{2}}(\tau)
+ B_N M_{\kappa_N , \frac{1}{2}}(\tau),~\tau = \nu_N y
\end{equation}
can be written in terms of the Whittaker functions
$W_{\kappa_N , \frac{1}{2}}~\mbox{and}~M_{\kappa_N , \frac{1}{2}}$ \cite{Abrsteg}.
In the "electric" region
eq. (\ref{E:main}) reads

\begin{equation}\label{E:eleq}
\xi_N^{''}(x) - G_N (x)\xi_N(x)= 0,
\end{equation}
where

$$
G_N (x) = x +
\frac{2f_N ^{-\frac{1}{3}}}{x-\kappa_N^{-2}f_N ^{-\frac{2}{3}}};~
x=-\frac{y}{b_N} + \kappa_N^{-2} f_N ^{-\frac{2}{3}};~
b_N (F) = \left(\frac{\hbar^2}{2\mu_N eF}  \right)^{\frac{1}{3}}.
$$
Following the comparison equation method \cite{Slav} the general solution normalized according to $\delta (W_N -W_N')$

\begin{equation}\label{E:elsol}
\xi_N(y) = \tilde{C}_N
\frac{\left[\frac{3}{2}S_N (x)  \right] ^{\frac{1}{6}}}{G_N(x)^{\frac{1}{4}}}
\left\{\sin\vartheta_N
Ai\left[\left(\frac{3}{2}S_N (x)  \right)^{\frac{2}{3}}  \right]+
\cos\vartheta_N Bi\left[\left(\frac{3}{2}S_N (x)  \right)^{\frac{2}{3}}  \right]  \right\}
\end{equation}
is written in terms of the Airy functions $Ai$ and $Bi$ \cite{Abrsteg}.
In eq. (\ref{E:elsol})

$$
S_N (x) = \int_0^x G_N^{\frac{1}{2}}(u)du,\quad\tilde{C}_N^2 =
\frac{1}{\mathscr{E}_N b_N},\quad
\mathscr{E}_N(F)=\frac{\hbar^2}{2\mu_N b_N^2},
$$
and $\vartheta_N$ is an arbitrary phase.

A comparison of function (\ref{E:iter}) and (\ref{E:coulomb})
for $\tau \ll 1$ and then the asymptotic expansions of functions
(\ref{E:coulomb}) for $\tau \gg 1$ and (\ref{E:elsol})
for $S \gg 1$ \cite{Abrsteg} lead to the set of equations

\begin{equation}\label{E:set11}
c_N -\frac{A_N}{\Gamma (1-\kappa_N)}=0;
\end{equation}

\begin{equation}\label{E:set12}
A_N Y_N(\kappa_N) +B_N\Gamma (-\kappa_N)=0;
\end{equation}

\begin{equation}\label{E:set13}
A_N -B_N\frac{\cos\pi \kappa_N }{\Gamma (1+\kappa_N)}-
\tilde{C}_N\pi^{-\frac{1}{2}}\kappa_N^{\frac{1}{2}}
f_N^{\frac{1}{6}}\cos\vartheta_N \Omega_N^{-1}=0;
\end{equation}

\begin{equation}\label{E:set14}
B_N\frac{1}{\Gamma (1-\kappa_N)}-
\tilde{C}_N\frac{1}{2}\pi^{-\frac{1}{2}}\kappa_N^{\frac{1}{2}}
f_N^{\frac{1}{6}}\sin\vartheta_N \Omega_N=0.
\end{equation}
In this set

\begin{equation}\label{E:Y}
Y_N(\kappa_N)=\pi\cot\pi\kappa_N - \frac{1}{2\kappa_N} -
\ln\kappa_N +\psi(1+ \kappa_N) +\ln q +
\ln \frac{\mid N -\tilde{\sigma} \mid \pi}{2} +2C-\frac{1}{2},
\end{equation}
where $C=0.577$ is the Euler constant,
$\psi(1+ \kappa_N)$ is the logarithmic derivative of the $\Gamma$
function,

\begin{equation}\label{E:exp}
\Omega_N = \exp\left\{-\frac{2}{3f_N\kappa_N^3}+
\kappa_N \ln\frac{8}{f_N\kappa_N^3}  \right\}.
\end{equation}
Solving the set of equations (\ref{E:set11})-(\ref{E:set14}) with respect to
the coefficients $c_N, A_N, B_N, \tilde{C}_N$
by the determinantal method, we determine the phase $\vartheta_N$
and then the coefficient $c_N^2 = \xi_N^2(0)$

\begin{equation}\label{E:coeff}
c_N^2= \frac{1}{2\pi p q}
\frac{\kappa_N \Gamma^2 (-\kappa_N)}{Y^2(\kappa_N)}\Omega_N^2
\left\{1 + \frac{\Omega_N^4\Gamma^4 (1-\kappa_N)}{4\kappa_N^2}
\left[\frac{1}{Y_N(\kappa_N)} -\frac{\sin 2\pi\kappa_N}{2\pi}  \right]^2\right\}^{-1}
\end{equation}
which provides us with the absorption coefficient  (\ref{E:absgen})
$\alpha_N=\alpha^{(0)}\frac{2p}{n_b\mid N -\tilde{\sigma} \mid}c_N^2$.
The obtained results are valid under the condition
~$f_N\kappa_N^3 \ll 1$, resulting in ~$\Omega_N \ll 1$.
\\
\\
\newpage

 \emph{\textbf{3.1.1 Exciton electroabsorption}}
\\
\\
Expanding function (\ref{E:Y}) in the vicinity of the discrete exciton energies,
specified by the quantum numbers $\kappa_{Nn}$
calculated from equation

\begin{equation}\label{E:discen}
Y_N(\kappa_{Nn})=0;~\kappa_{Nn} = n + \beta_{Nn};~ \beta_{Nn} < 1;~ n=0,1,2,\ldots ,
\end{equation}
eq. (\ref{E:coeff})
is rearranged to

\begin{eqnarray}\label{E:peak}
c_{Nn}^2=\frac{q\Delta_N}
{4p\kappa_{Nn}^3\mid\frac{\partial Y_N}{\partial\kappa_{Nn} }\mid }
\Lambda_{Nn}\left( W_N  \right);
\quad \frac{\partial Y_N
}{\partial\kappa_{Nn}}=\left\{
\begin{array}{ll}
- (2\kappa_{Nn}^2)^{-1},\, n=0;\, \kappa_{Nn} = \beta_{N0};\\
 - \beta_{Nn}^{-2},\, n=1,2,\ldots;\,\kappa_{Nn} = n + \beta_{Nn},
\end{array}
\right.
\end{eqnarray}
where the function $\Lambda_{Nn}\left( W_N  \right)$
describing the Lorentzian form of the optical absorption peak reads

\begin{equation}\label{E:lor}
\Lambda_{Nn}(W_N)=\frac{\Gamma_{Nn}^{\left(\mbox{\tiny{el}}\right)}}
{2\pi
\left\{\left[W_N -W_{Nn} -\Delta W_{Nn}^{(r)} \right]^2 +
\frac{\Gamma_{Nn}^{\left(\mbox{\tiny {el}}\right)2}}{4} \right\}};~W_N = \hbar \omega - \Delta_N.
\end{equation}
In eq. (\ref{E:lor}) $W_{Nn} = -Ry_N(n + \beta_{Nn})^{-2},
\quad \Gamma_{Nn}^{\left(\mbox{\tiny{el}}\right)}~\mbox{and}~\Delta W_{Nn}^{(r)}\sim \Gamma_{Nn}^{\left(\mbox{\tiny {el}}\right)}\Omega_N^2 \beta_{Nn}^{-1}\ll \Gamma_{Nn}^{\left(\mbox{\tiny {el}}\right)}$
are the exciton energy levels in the absence of the electric field
counted from the energy gap $\Delta_N$ and the resonant widths and shifts of the
optical peaks, respectively, caused by the electric field ionization
of the exciton states. We have omitted here the negligibly small Stark
corrections to the energies $W_{Nn}$ caused by the weak electric fields (see Sec. 5, subsec. 5.1.1)
and further ignore the resonant
shifts $\Delta W_{Nn}^{(r)}$. The reasons for the latter
approximation
are as follows: (i) the shifts are much smaller than the corresponding widths
and (ii) they do not change the form
of the exciton spectrum.
The widths $\Gamma_{Nn}^{\left(\mbox{\tiny {el}}\right)}$ transform the $\delta$-function
peaks in
an unbiased AGNR \cite{Monschm16} into the maxima
of finite width $\Gamma_{Nn}^{\left(\mbox{\tiny {el}}\right)}$.

An explicit form of the width $\Gamma_{Nn}^{\left(\mbox{\tiny {el}}\right)}$ and maximum
of the absorption coefficient $c_{Nn}^{2(\mbox{\tiny{max}})}$
determined by eqs. (\ref{E:coeff}) and (\ref{E:peak}),
respectively, are as follows

\begin{equation}\label{E:grexc}
\Gamma_{Nn}^{\left(\mbox{\tiny {el}}\right)} = 2 (1 + \delta_{n0})\frac{Ry_N \Omega_N^2}{\kappa_{Nn}^2\Gamma^2 (1+ \kappa_{Nn} ) },\quad
c_{Nn}^{2(\mbox{\tiny{max}})} =
\frac{2\beta_{Nn}^2\Gamma^2 (1+ \kappa_{Nn} ) }{\pi p q \kappa_{Nn}\Omega_N^2  },
\end{equation}
where $\delta_{n0}$ is the Kronecker symbol. The quantum defects
$\beta_{Nn}~\mbox{and factor}~\Omega_N$ can be found from eqs.
(\ref{E:discen}) and (\ref{E:exp}), respectively. As expected,
in the absence of the electric fields $F=0,~\Omega_N = 0$
eqs. (\ref{E:peak}) and (\ref{E:lor}) transform into those describing
the $\delta$-function type exciton peaks
in an electrically unbiased AGNR \cite{Monschm16}.
In principle,
the resonant widths $\Gamma_{Nn}^{\left(\mbox{\tiny {el}}\right)}$ and shifts
$\Delta W_{Nn} \sim \frac{\Omega_{Nn}^2}{\beta_{Nn}}\Gamma_{Nn}^{\left(\mbox{\tiny {el}}\right)}$
of the exciton peaks can be found as the imaginary components
of the complex energies of the
quasi-discrete $Nn$ states \cite{Monschm14}.
These states are relevant to
the poles
of the scattering matrix
$S(\vartheta_N)= \exp (2{\rm i}\vartheta_N )$ \cite{Newt}.
Setting in the set of eqs. (\ref{E:set11})-(\ref{E:set14})
$\cot \vartheta_N = {\rm i}$ and then solving this
set by the determinantal method we find the complex quantum
numbers $\kappa_{Nn}$ and the complex energies
$W_N = W_{Nn} + \Delta W_{Nn} - \frac{{\rm i}}{2}\Gamma_{Nn}^{\left(\mbox{\tiny {el}}\right)}$
comprising the unperturbed Rydberg levels $W_{Nn}=-Ry_N/\kappa_{Nn}^2$, resonant width $\Gamma_{Nn}^{\left(\mbox{\tiny {el}}\right)}$ (\ref{E:grexc}) and shift $\Delta W_{Nn} \sim \frac{\Omega_{Nn}^2}{\delta_{Nn}}\Gamma_{Nn}^{\left(\mbox{\tiny {el}}\right)}$.
\\
\\
 \emph{\textbf{3.1.2 Franz-Keldysh exciton absorption}}
\\
\\
Here we consider the frequencies $\hbar\omega < \Delta_N$,
positioned away from the quasi-discrete exciton
$Nn$ peaks. It allows us
to neglect in eq. (\ref{E:coeff}) the small term
$\sim \Omega_N^4 \Gamma^4 (1-\kappa_{N}) \kappa_{N}^{-2}$.
This equation becomes

\begin{equation}\label{E:excfrk}
c_N^2=c_{N(\mbox{\tiny {F-K}})}^2T_N
\end{equation},
where

\begin{equation}\label{E:frkeld}
c_{N(\mbox{\tiny{F-K}})}^2=\frac{1}{\sqrt{2}\pi p}
\left(\frac{\Delta_N - \hbar \omega}{\Delta_N}\right)^{-\frac{1}{2}}
\exp\left\{-\frac{4}{3}\left(\frac{1}{f_N \kappa_{N}^3} \right)^{\frac{3}{2}}\right\}
\end{equation}
is the electroabsorption coefficient in the AGNR
associated with the photon-assisted inter-subband tunneling of the
free carriers (F-K effect \cite{Frkeld})
and

\begin{equation}\label{E:exceff}
T_N=\frac{\Gamma^2(-\kappa_{N})}{4Y_N^2(\kappa_{N})}
\exp\left\{2\kappa_{N}\ln \frac{8}{f_N\kappa_{N}^3} \right\}
\end{equation}
is the factor
describing the exciton
influence on the F-K absorption.
The equations (\ref{E:excfrk}-\ref{E:exceff}) are derived
for the frequency shift $\Delta_N - \hbar\omega$ significantly
exceeding the energy  $\mathscr {E}_N\,(f_N\kappa_{N}^3 \ll 1) $.
As expected, in the absence of the exciton effect $(\kappa_{N} \rightarrow 0)\, ~\Gamma(-\kappa_{N})^2 = 4 Y(\kappa_{N}^2),~
T_N(\kappa_{N})=1\,\mbox{and}\,c_N^2 = c_{N(\mbox{\tiny {F-K}})}^2$.
For the zeroth electric fields $F=0$ the coefficient $c_{N(\mbox{\tiny {F-K}})}^2=0$
which in turn leads to the zeroth absorption $c_N^2 =0$ for the frequencies distant from the exciton peaks.

\subsection{Frequency region $\hbar \omega - \Delta_N > 0 $}

For these energies
we introduce the quantum number $\zeta_N$ and parameter $s_N$
defined by $W_N=Ry_N \zeta^{-2}$ and
$s_N = 2(\zeta_N a_N)^{-1}$, respectively. In the Coulomb region
the general solution to eq. (\ref{E:main}) has the form

\begin{equation}\label{E:coulomb1}
\xi_N (y)=D_N\left[e^{{\rm i}\Theta_N}W_{{\rm i} \zeta_N ,
\frac{1}{2}}(t)+  e^{-{\rm i}\Theta_N}W_{-{\rm i}\zeta_N,
\frac{1}{2}}(-t)\right];~t=-{\rm i}s_N y,
\end{equation}
where $\Theta_N$ is an arbitrary phase.

In the "electric" region
equation (\ref{E:main}) reads

\begin{equation}\label{E:eleq1}
\xi_N^{''}(x) - \tilde{G}_N (x)\xi_N(x)= 0,
\end{equation}
where

$$
\tilde{G}_N (x) = x +
\frac{2f_N^{-\frac{1}{3}}}{x+\zeta_N^{-2} f_N ^{-\frac{2}{3}}};~
x=-\frac{y}{b_N} - \zeta_N^{-2} f_N ^{-\frac{2}{3}}
$$
with the general solution

\begin{equation}\label{E:elsol1}
\xi_N(y) = \tilde{C}_N
\frac{\left[\frac{3}{2}\tilde{S}_N (x)\right]^{\frac{1}{6}}}{\tilde{G}_N(x)^{\frac{1}{4}}}
Ai\left[\left(\frac{3}{2}\tilde{S}_N (x)  \right)^{\frac{2}{3}}  \right],
\end{equation}
decreasing towards the region $y \ll y_N~(y_N <0)$
and normalized according to $\delta \left(W_N -W_N^{'}\right)$.
In eq. (\ref{E:elsol1})

$$
\tilde{S}_N (x)=\int_0^x \tilde{G}_N (u)du
$$
and $\tilde{C}_N$ is the same as that in eq. (\ref{E:elsol}).

A comparison of functions
(\ref{E:iter}) and (\ref{E:coulomb1}) within the inner region
and then the asymptotic
expansions of functions (\ref{E:coulomb1}) and (\ref{E:elsol1})
for $\mid t \mid \gg 1$ and $\tilde{S}_N \gg 1$, respectively,
lead to the set of equations

\begin{equation}\label{E:contin}
c_N = -2D_N \left( \frac{\sinh \pi \zeta_N}{\pi \zeta_N} \right)^{\frac{1}{2}}
\sin\left(\Theta_N + \sigma_N \right),\quad \sigma_N = \mbox{arg} \Gamma
({\rm i} \zeta_N),
\end{equation}

\begin{equation}\label{E:coeffs}
2D_N{\rm e}^{\frac{\pi\zeta_N}{2}}=
\tilde{C}_N \pi^{-\frac{1}{2}} \zeta_N^{\frac{1}{2}} f_N^{\frac{1}{6}},
\end{equation}

\begin{equation}\label{E:phase1}
\Theta_N = \frac{2}{3f_N \zeta_N^3}
-\zeta_N \ln \frac{8}{f_N \zeta_N^3} -\frac{\pi}{4}.
\end{equation}
This set results in the coefficient
$c_N =\xi_N (0)$ with

\begin{equation}\label{E:excfrkcont}
c_N^2 = \frac{1}{\pi p}
\left(\frac{\Delta_N}{\hbar \omega - \Delta_N}\right)^{\frac{1}{2}}
Z_N \sin^2 \left(\frac{2}{3f_N \zeta_N^3} + \frac{\pi}{4} + \Delta \Theta_{NZ} \right)
\end{equation}
totally determining the absorption coefficient  (\ref{E:absgen})
$\alpha_N=\alpha^{(0)}\frac{2p}{n_b\mid N -\tilde{\sigma} \mid}c_N^2$
in the frequency region $\hbar \omega - \Delta_N > 0$.

The Sommerfeld factor $Z_N$ and phase shift $\Delta \Theta_{NZ}$
both responsible for the influence of the exciton on the
F-K absorption associated with the unbound electron-hole pair
are given by

\begin{equation}\label{E:Somm}
Z_N ={ \rm e}^{-\pi\zeta_N}\frac{\sinh \pi \zeta_N}{\pi \zeta_N}
~\mbox{and}~\Delta \Theta_{NZ} =
-\zeta_N \ln \frac{8}{f_N \zeta_N^3} -\frac{\pi}{2} + \sigma_N,
\end{equation}
respectively. Eqs. (\ref{E:excfrkcont}) and (\ref{E:Somm}) are
valid for frequency shifts from the edge by an amount
$\hbar \omega - \Delta_N$ considerably exceeding the energy
$\mathscr{E}\,(f_N \zeta_N^3\ll 1)$. In the absence of the
exciton effect $(\zeta_N \rightarrow 0)$ we arrive at the Sommerfeld factor $Z_N = 1$ and phase shift  $\Delta \Theta_{NZ}= -\pi$. This
transforms eq. (\ref{E:excfrkcont}) into that for the F-K oscillations. In the case of a vanishing electric field $F=0$
the rapidly oscillating function transforms into 1/2 and we obtain
the coefficient $c_N^2$ in eq. (\ref{E:excfrkcont}) for the exciton
absorption in the continuous spectrum region \cite{Monschm16}.

\section{Spectrum of the exciton electroabsorption:
Double-subband approximation}\label{S:spectr1}

In this section we consider the influence of the coupling between the
continuous and discrete states emanating from the
ground $\Delta_0$ and adjacent to the first excited
$\Delta_1$ size-quantized energy levels, respectively.
We focus on the exciton electroabsorbtion
in the frequency region $\Delta_0 < \hbar \omega < \Delta_1$
in the vicinity of the resonant peaks
$\hbar \omega - \Delta_1 = - Ry_1/\kappa_{1n}^2.$
The common resonant energy $E=\hbar \omega$ of the interacting states is

\begin{equation}\label{E:reson}
E=\Delta_1 - \frac{Ry_1}{\kappa_1^2}=
\Delta_0 +  \frac{Ry_0}{\zeta_0^2}.
\end{equation}

The corresponding two-fold set of equations follows from the general
one (\ref{E:main}) limited to $N,N' =0,1$. To avoid
routine and cumbersome calculations
only an outline of the mathematical procedure will be given below.
The needed details can be found in works in which the problems
of the exciton electroabsorption in QWR \cite{Monschm09} and  impurity states in
electrically biased AGNR \cite{Monschm14} have been studied. Using the trial functions and its derivatives
$\xi_1^{(0)}(y)=c_1, \xi_0^{(0)}(y)=c_0, \xi_1^{(0)'}(y)=0,
\xi_0^{(0)'}(y)=0$
and double integrating the two-fold set
we arrive at the functions
$\xi_1(y)~\mbox{and}~\xi_0(y)$ valid in the inner
regions $d \ll y \ll \kappa_1 a_1~\mbox{and}~
d \ll y \ll  \zeta_0 a_0$.
Matching these functions
with the functions (\ref{E:coulomb}) for $\tau \ll 1, N=1$
and (\ref{E:coulomb1}) for $\mid t \mid\ \ll 1, N=0$,
respectively, we obtain

\begin{equation}\label{E:ddiscr1}
\frac{A_1}{\Gamma(-\kappa_1)} + \kappa_1 c_1=0,
\end{equation}

\begin{equation}\label{E:ddiscr2}
\frac{A_1}{\Gamma(-\kappa_1)}Y_1(\kappa_1) +B_1 -\kappa_1\gamma_{01}c_0=0,
\end{equation}

\begin{equation}\label{E:dcont1}
c_0 + 2D_0\left( \frac{\sinh \pi \zeta_0}{\pi \zeta_0} \right)^{\frac{1}{2}}
\sin\left( \Theta_0 + \sigma_0  \right)=0,
\end{equation}

\begin{equation}\label{E:dcont2}
c_0 \eta_0(\zeta_0) + \gamma_{01} c_1 =0.
\end{equation}
In these equations

\begin{eqnarray}
&&\eta_0(\zeta_0)=\lambda_0 (\zeta_0) -
 \frac{\pi}{1-\rm e^{-2\pi\zeta_0 }}\cot \left( \Theta_0 + \sigma_0 \right);\quad
 \sigma_0 = \mbox{arg}\Gamma({\rm i}\zeta_0);
 \nonumber \\
&&\lambda_0 (\zeta_0) = \frac{1}{2}[\psi (1+{\rm i}\zeta_0) + \psi (1-{\rm i}\zeta_0)]
-\ln \zeta_0 + \ln \frac{\mid \tilde{\sigma}  \mid \pi}{2}
+2C - \frac{1}{2} +\ln q.
\end{eqnarray}
The parameter

$$
\gamma_{01} = \frac{1}{2d^2}\int_{-\frac{d}{2}}^{\frac{d}{2}}dx_e
\int_{-\frac{d}{2}}^{\frac{d}{2}}dx_h  \ln \mid x_e -x_h \mid
\sin\frac{\pi}{d}x_e \sin\frac{\pi}{d}x_h = 0.387
$$
describes the coupling induced by the off-diagonal potentials
$V_{10}=V_{01}$ (\ref{E:1Dc}). The function $Y_1 (\kappa_1)$ is given by eq. (\ref{E:Y}) for $N=1$.

Comparing functions (\ref{E:coulomb})
for $\tau \gg 1$ and (\ref{E:elsol}) for $S\gg 1$ both for
$N=1$ we obtain equations (\ref{E:set13}) and (\ref{E:set14})
both for $N=1$. The relationship between the phases
$\vartheta_1$ and $\Theta_0$ is derived by equating
the phases in the asymptotic expansions of functions
(\ref{E:elsol}) for $S\gg 1$ and (\ref{E:coulomb1})
for $\mid t \mid \gg 1$ to give

\begin{equation}\label{E:phase2}
\Theta_0 + \sigma_0 = - \vartheta_1 + \frac{2}{3f_1 \kappa_1^3}
-\kappa_1 \ln \frac{8}{f_1 \kappa_1^3} -\frac{\pi}{4}.
\end{equation}

Solving the set of eqs. (\ref{E:ddiscr1})-(\ref{E:dcont2})
and (\ref{E:set13}), (\ref{E:set14}) for $N=1$ with respect to
the coefficients $c_0, c_1, A_1, B_1, D_0, \tilde{C}_1$
by the determinantal method, we obtain the equation containing the phase $\vartheta_1$ and $\Theta_0 + \sigma_0$ linked by eq. (\ref{E:phase2})

\begin{equation}\label{E:detd}
\eta_0(\Theta_0)\left\{Y_1(\kappa_1)-
\left[\frac{\sin 2\pi\kappa_1 }{2\pi} +
\frac{2\cot\vartheta_1}
{\kappa_1\Gamma^2(-\kappa_1)\Omega_1^2}\right]^{-1}\right\}
-\gamma_{01}^2 =0.
\end{equation}

As expected, the just derived set of equations and eq. (\ref{E:detd}) satisfy the limiting transitions.
In the single subband approximations neglecting the coupling
parameter $\gamma_{01}$, eq. (\ref{E:detd}) decomposes into two ones. Determining $\cot\vartheta_1 $, by setting to zero the curly bracket we arrive at the absorption coefficient
$\alpha_1 \sim c_1^2$ (\ref{E:coeff}) describing the exciton
electroabsorption  in the frequency region
$\hbar \omega < \Delta_1$. For a vanishing electric field
$F=0,~ \Omega_1 = 0$ eq. (\ref{E:detd}) generates the equation
$\eta_0=\gamma_{01}^2 Y_1^{-1}$ that in turn allows us
to determine $\cot(\Theta_0 + \sigma_0)$. Using eqs.
(\ref{E:ddiscr1})-(\ref{E:dcont2}) with
$B_1 =0$ and $D_0 = (4\pi p q)^{-1/2}\zeta_0^{1/2}\rm e^{-\pi\zeta_0/2 }$
providing the normalization of function (\ref{E:coulomb1})
to $\delta(W_0 -W_0')$ we arrive to the absorption coefficient
$\alpha_1 \sim (c_1+ c_0)^2$. This coefficient is determined
by eqs. (\ref{E:peak}) and (\ref{E:lor}) in which
the resonant width $\Gamma_{1n}^{\left(\mbox{\tiny{el}}\right)}$
and corresponding resonant shift are replaced by the Fano resonant
width $\Gamma_{1n}^{\left(\mbox{\tiny{F}}\right)}$ and shift
$\sim q^3 \Gamma_{1n}^{\left(\mbox{\tiny{F}}\right)} $.
The latter have been calculated in Ref. \cite{Monschm16}
to give

\begin{equation}\label{E:fano}
\Gamma_{1n}^{\left(\mbox{\tiny{F}}\right)}=
\frac{8 q\gamma_{01}^2Ry_1}{\sqrt{3}\kappa_{1n}^3 \mid \frac{\partial Y_1}{\partial \kappa_1}   \mid    }.
\end{equation}
Note that this resonant width can be calculated as the imaginary
part of the complex energy determined by the complex quantum
number $\kappa_{Nn}$ in solving eqs.
(\ref{E:ddiscr1}), (\ref{E:ddiscr2}) for $B_1 = 0$ and (\ref{E:dcont2})
for $\cot (\Theta_0 + \sigma_0)=\rm{i}$ by the determinantal method.

Here we consider the case of the resonant state $1n$ for which both ionization channels are open. In view of
$\zeta_0 \simeq q/2\sqrt{2}\ll 1$ (see eq. (\ref{E:reson})) and the resonant state condition
$\cot\vartheta_1 = \rm{i}$ \cite{Newt} we calculate
from eq. (\ref{E:phase2}) $\cot(\Theta_0 + \sigma_0)=\rm{i}$ and then from eq. (\ref{E:detd}))
the complex quantum number $\kappa_{1n}^*$. The imaginary part of the energy $W_{1n}(\kappa_{1n}^*)$
leads to the energetic width $\Gamma_{1n}^{\left(\mbox{\tiny{t}}\right)}$ of the
corresponding quasi-discrete state with

\begin{equation}\label{E:total}
\Gamma_{1n}^{\left(\mbox{\tiny{t}}\right)} =
\Gamma_{1n}^{\left(\mbox{\tiny{el}}\right)} + \Gamma_{1n}^{\left(\mbox{\tiny{F}}\right)},
\end{equation}
where the widths $\Gamma_{1n}^{\left(\mbox{\tiny{el}}\right)}$
and $\Gamma_{1n}^{\left(\mbox{\tiny{F}}\right)}$ are determined by equations
(\ref{E:grexc}) and (\ref{E:fano}), respectively.
Now in view of
$\cot\vartheta_1 = \cot(\Theta_0 + \sigma_0)$
we find from eqs. (\ref{E:detd})
and from the set of eqs.
(\ref{E:ddiscr1})-(\ref{E:dcont2}),
(\ref{E:set13}), (\ref{E:set14}) for $N=1$ the absorption coefficient
$\alpha_1 \sim (c_1+ c_0)^2$. In the vicinity of the
resonant states specified
by the quantum numbers $\kappa_{1n}^*$
the absorption
coefficient acquires the forms
(\ref{E:peak}) and (\ref{E:lor}), in which
the width
$\Gamma_{1n}^{\left(\mbox{\tiny{el}}\right)}$ is replaced, as expected, by the total width
$\Gamma_{1n}^{\left(\mbox{\tiny{t}}\right)}$
from eq. (\ref{E:total}).

\section{Discussion}\label{S:disc}

In this section we discuss the exciton electroabsorption spectrum
obtained both for the isolated subbands i.e. the exciton peaks (5.1.1) and continuous Franz-Keldysh exciton absorption (5.1.2, 5.2)  and in
view of intersubband Fano coupling (5.3). We focus on the
dependencies of the spectral characteristics on the ribbon width
and electric field strength.
Estimates of the expected
experimental values and comparison of our results with those obtained numerically by other authors are given.
\\
\\
\textbf{5.1 Single subband approximation}
\\
\\
In the single subband approximation the electroabsorption spectrum
consists of a periodic
sequence of $N$ subbands each comprising
the Rydberg series of the quasi-discrete peaks
adjacent to the edges $\Delta_N$ from below.
In addition, there is continuous bands covering the
spectral regions both below
and above the edges.
The intensities
of the quasi-discrete peaks, their electrically induced widths and the shape of the continuous
band depend on both the ribbon width $d$
and electric field strength $F$.
The peak positions are predominantly determined
by the ribbon width.
\\
\\
\textbf{5.1.1 Exciton electroabsorbtion $\hbar\omega < \Delta_N$}
\\
\\
Here we address the Rydberg series of spectral maxima
described by eqs. (\ref{E:peak}) and (\ref{E:lor})
in which we ignore both the Stark $\Delta W_{Nn}^{(S)}$
and the resonant tunneling $\Delta W_{Nn}^{(r)}$ shifts
of the energy level $W_{Nn}$. The reason for this is that
for weak electric fields $F$ for which
$f_N \kappa_{Nn}^3 \ll 1$ the contributions of both shifts
to the peak positions are negligibly small.
In order to justify
this we set in eq. (\ref{E:main}) the corrections
$\Delta \xi_{N0}\sim f_N $ to the ground state $n=0$ wave function
$\xi_{N0} \sim \exp (- \tau/2) $ and to the energy $\Delta W_{N0}^{(S)} \sim -f_N^2$ (red shift).
We obtain
$\mid\Delta W_{N0}^{(S)}\mid/W_{N0}  =(5/4)f_N^2\kappa_{N0}^6 \ll 1$.
This result
coincides with that
calculated by Ratnikov and Silin \cite{Rat} by the
Dalgarno-Lewis perturbation theory method \cite{Dalg}.
For the excited exciton peaks the electric field effect
is equally small. The resonant shift $W_{Nn}^{(r)}$
in eq. (\ref{E:lor}) calculated from eq. (\ref{E:coeff})
becomes $\Delta W_{Nn}^{(r)}\sim \Gamma_{Nn}^{\left(\mbox{\tiny{el}}\right)}
\left(\Omega_N^2/\beta_{Nn} \right)$. It turns out to be not only much smaller than the energy
$W_{Nn}$ but also width $\Gamma_{Nn}^{\left(\mbox{\tiny{el}}\right)}$.

The dependence of the
binding
energies $E_{Nn}^{\mbox{\tiny (b)}}= -W_{Nn} \sim \Delta_N~\mbox{with}~
\Delta_N \sim d^{-1} $ and exciton peak positions
$\hbar\omega_{Nn}= \Delta_N - E_{Nn}^{\mbox{\tiny (b)}}$ on the ribbon width $d$
is the same as that in the unbiased
AGNR \cite{Monschm16}. The narrowing ribbon leads to an increasing binding energy and blueshift
of the exciton peak.

While it does not significantly change the peak positions, the electric field
modifies drastically the shape of the absorption peaks. The
$\delta$-function form for the unbiased ribbon \cite{Monschm16}
is replaced by a Lorentzian one determined by the width
$\Gamma_{Nn}^{\left(\mbox{\tiny{el}}\right)}$ and finite absorption intensity maximum $\alpha_{Nn}^{\mbox{\tiny (max)}}$,
derived from eqs. (\ref{E:absgen}) and (\ref{E:grexc})

\begin{equation}\label{E:max}
\alpha_{Nn}^{\mbox{\tiny (max)}} = \alpha^{(0)}
\frac{8}{n_b \mid  N-\tilde{\sigma}\mid \pi q}
\frac{\beta_{Nn}^2 \Gamma^2 (1+\kappa_{Nn} )}
{\kappa_{Nn}\Omega_{Nn}^2};~n=0,1,2,\ldots .
\end{equation}
It follows from eqs. (\ref{E:grexc}) and (\ref{E:max})
that with increasing the ribbon width $d$ and
electric field strength $F$ the width
$\Gamma_{Nn}^{\left(\mbox{\tiny{el}}\right)}$ increases,
while the exciton absorption peak maximum
$\alpha_{Nn}^{\mbox{\tiny (max)}}$ decreases in magnitude.
Thus, in contrast to the quasi-1D semiconductor
structures (QWR, DE) in which the exciton
ionization is provided only by the
electric field, in the AGNR a dimensional ionization
can be realized. The ribbon widening leads
to the ionization, while the electric field
remains constant. The rate of the dimensional ionization
surpasses that of the electric field because
of the square
$\sim(d^{-1})^2$ and linear $\sim F^{-1} $ dependencies
of $\Omega_{Nn}$ (\ref{E:exp}) on the
reciprocal ribbon width $d$ and electric field strength $F$, respectively.
The width $\Gamma_{00}^{\left(\mbox{\tiny{el}}\right)}$
of the ground exciton state in the AGNR placed
on the sapphire substrate $(q=0.24)$ as a function
of the ribbon width $d$ and electric field strength $F$ is depicted
in Fig. 1. Isowidth lines $Fd^2 \simeq \,\mbox{const.} $
providing the constant width
$\Gamma_{00}^{\left(\mbox{\tiny{el}}\right)} (F,d)$
are given in Fig. 2.

\begin{figure}
    \centering
    \includegraphics[width=0.75\textwidth]{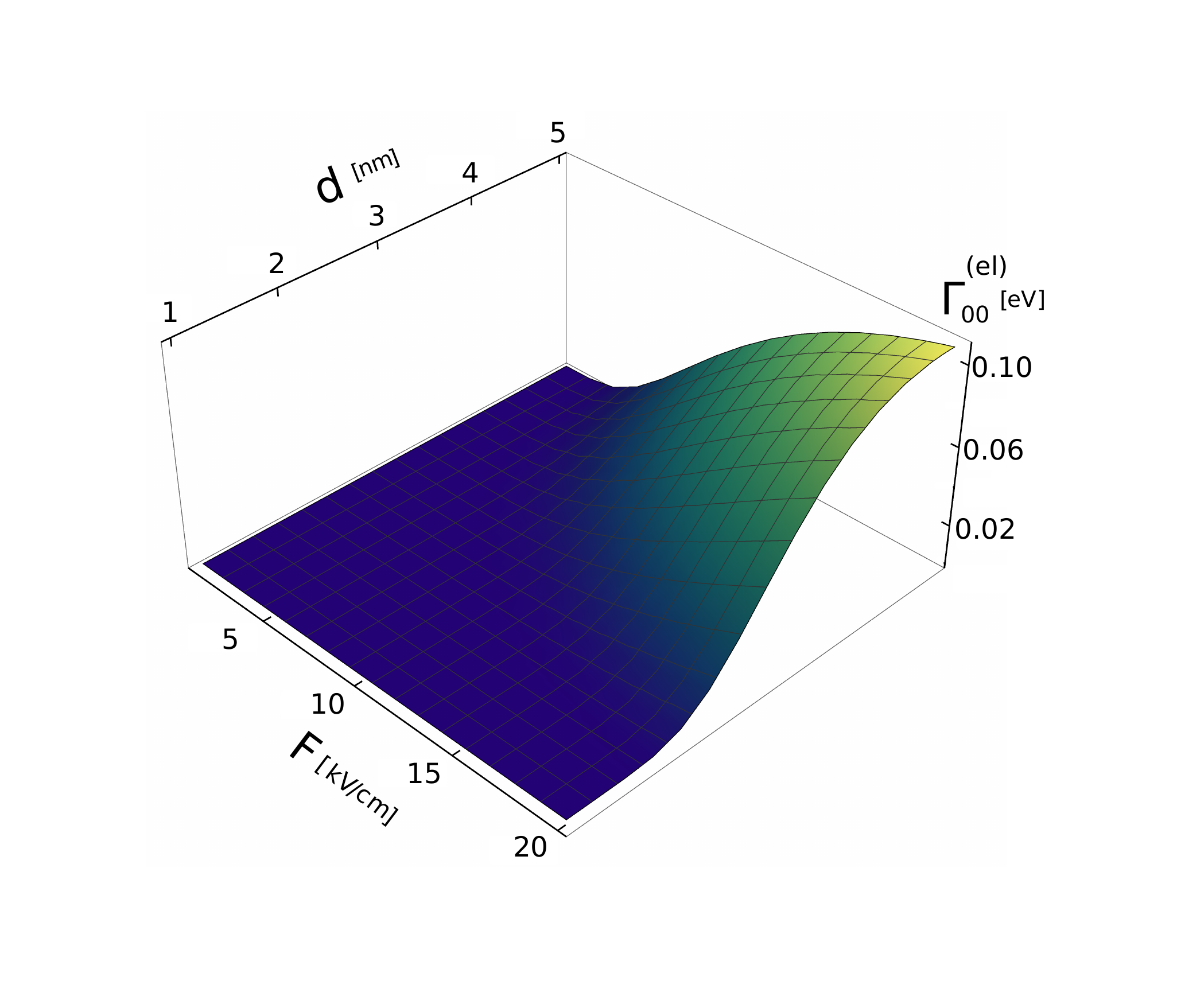}
    \caption{The width $\Gamma_{00}^{(\mbox{\tiny{el}})}$ of the ground exciton peak versus the ribbon width $d$ and electric field strength $F$. The width $\Gamma_{00}^{(\mbox{\tiny{el}})}$, quantum number $\kappa_{00}$ and factor $\Omega_{00}$ are determined via eqs. (\ref{E:grexc}), (\ref{E:discen}) and (\ref{E:exp}), respectively, for the ribbon placed on the sapphire substrate $q=0.24$.}
\end{figure}

\begin{figure}
    \centering
    \includegraphics[width=0.75\textwidth]{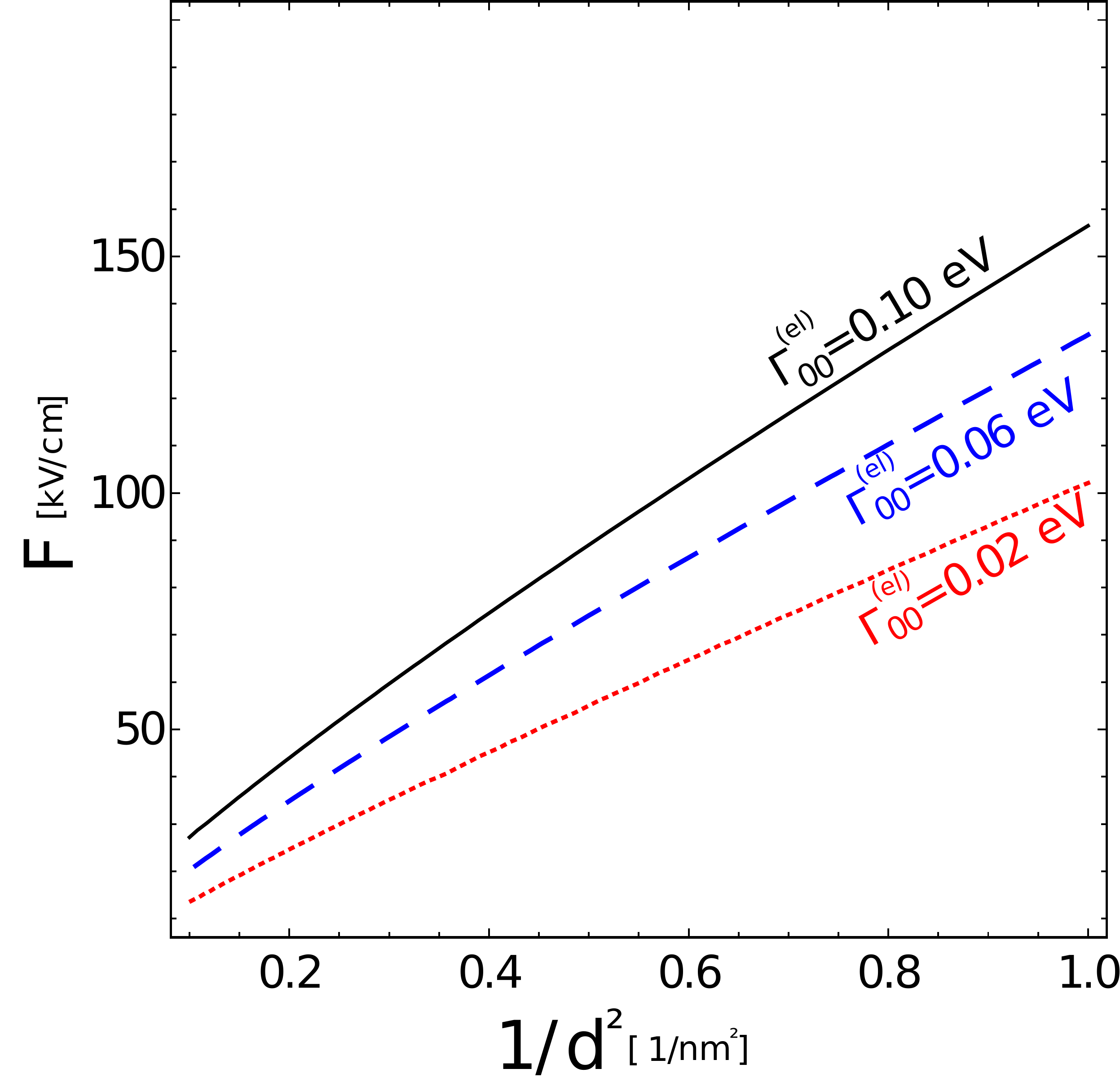}
    \caption{The isowidths curves $\Gamma_{00}^{(\mbox{\tiny{el}})}(d,F)=\mbox{const.}$ for the ground exciton peak. The width $\Gamma_{00}^{(\mbox{\tiny{el}})}$,
quantum number $\kappa_{00}$ and factor $\Omega_{00}$
are determined from the same equations and for the same ribbon substrate as those used in Fig.1. The values of  $\Gamma_{00}^{(\mbox{\tiny{el}})}$ are chosen to be
$0.10;~0.06;~0.02~(\mbox{meV})$.  }
\end{figure}

\begin{figure}
    \centering
    \includegraphics[width=0.75\textwidth]{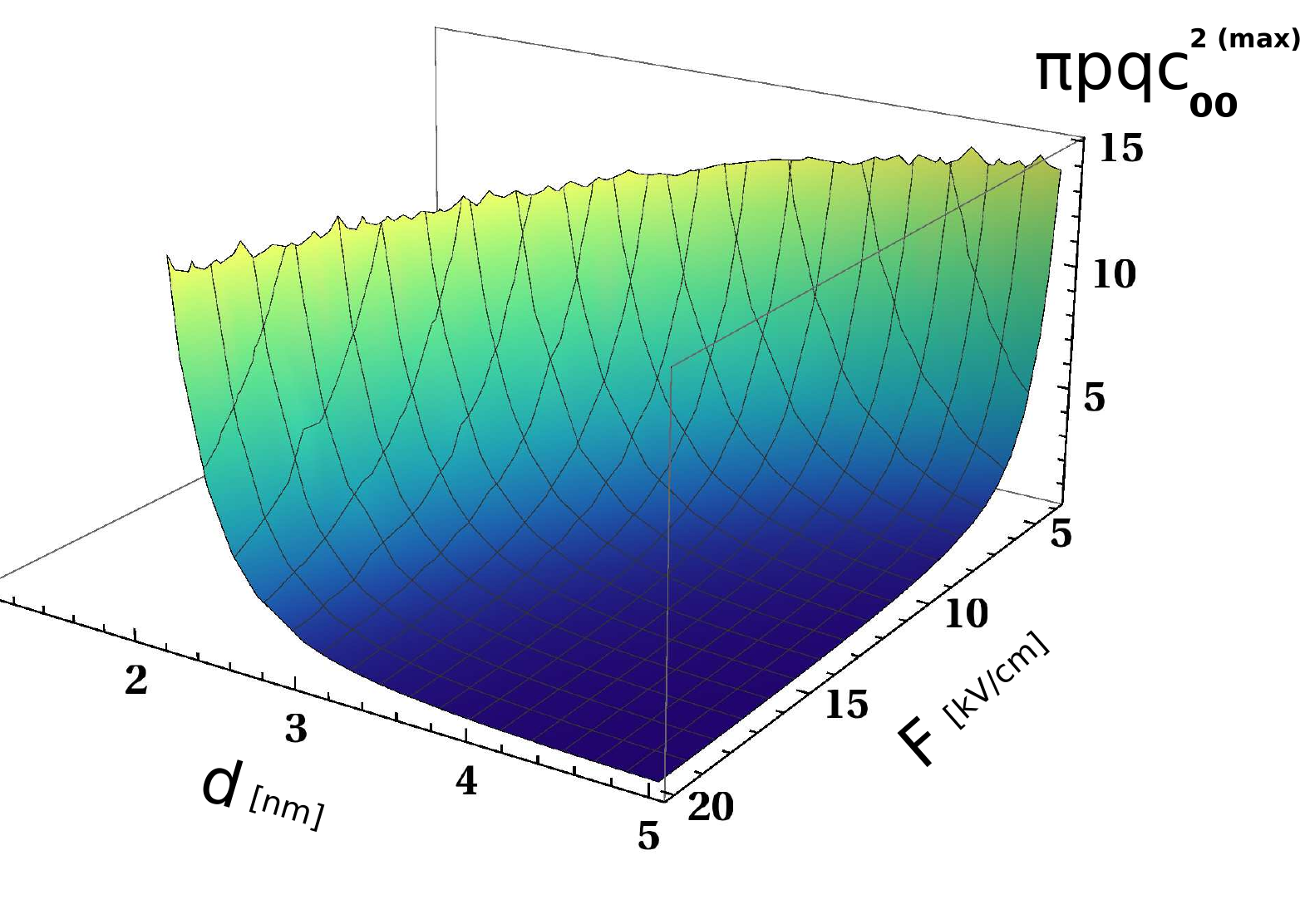}
    \caption{The dependence of the ground exciton peak maximum
$c_{00}^{2\mbox{\tiny(max)}}$ on the ribbon width $d$
and electric field strength $F$. The coefficient
$c_{00}^{2\mbox{\tiny(max)}}$,
quantum number $\kappa_{00}$ and factor $\Omega_{00}$
are determined from the same equations and for the same substrate as in Fig.1.}
\end{figure}

Eq. (\ref{E:grexc}) shows that within the $N$ Rydberg series
the lowest absorption line considerably surpasses
the excited peaks in narrowness
$\Gamma_{N0}^{\left(\mbox{\tiny{el}}\right)}
<<\Gamma_{Nn}^{\left(\mbox{\tiny{el}}\right)}$ and
intensity $\alpha_{N0}^{\mbox{\tiny (max)}}
>> \alpha_{Nn}^{\mbox{\tiny (max)}},~n=1,2,\ldots$.
Fig. 3 demonstrates the dependence of the ground exciton
peak maximum
$c_{00}^{2(\mbox{\tiny{max}})}$
on the ribbon width $d$ and electric field strength $F$.

Eq. (\ref{E:exp}) allows us to introduce the parameter
of the exciton state relative stability $Q_{Nn}$ in the ribbon with
width $d$ exposed to the electric field $F$. This parameter
generated by the condition $f_N\kappa_{Nn}^3 = 1$ becomes

\begin{equation}\label{E:stab}
Q_{Nn}(q)=\frac{\pi^2 p}{8e}\mid N-\tilde{\sigma}  \mid^2 \frac{q^3}{\kappa_{Nn}^3};\quad \frac{\pi^2 p}{8e}=0.82~\mbox{Vnm}
\end{equation}
where $\kappa_{Nn} (q)$ is the root of eq. (\ref{E:discen}).
Under the condiiton
$Fd^2 > Q_{Nn}(q)$ the $Nn$ state is practically ionized,
while in the opposite case $Fd^2 < Q_{Nn}(q)$
the state in question remains
relatively stable. An electric field providing the relatively small width of the ground peak $\kappa_{N0}< 1$
would be sufficiently strong to ionize
completely the excited
states $\kappa_{Nn} > n,\, n=1,2, \ldots$.
Note that the parameter $Q_{Nn}(q)$ possesses a universal character as far as it depends only on the dimensionless exciton potential strength $q$.
\\
\\
\textbf{5.1.2 Franz-Keldysh exciton absorption} $\hbar\omega < \Delta_N$
\\
\\
As pointed out originally by Merkulov and Perel \cite{Merper} the mechanism of the interband electroabsorption in bulk materials in the spectral regions distant from the exciton maxima is the F-K effect \cite{Frkeld} (i.e. optical transition assisted by the interband carriers tunneling) strongly modified by the exciton attraction.
This is completely
in line with eqs. (\ref{E:excfrk})-(\ref{E:exceff})
relevant to the exciton electroabsorption and F-K effect
in the AGNR. In the latter this spectrum
depends both on the electric fields $F$ and ribbon width $d$,
while in the semiconductor
structures, i.e. in bulk crystals as well as in those subject to strong magnetic fields
\cite{Mon76} and QWR
\cite{Monschm09} mostly only electric fields $F$ influence the optical absorption spectrum. It follows from eqs.
(\ref{E:excfrk})-(\ref{E:exceff}) in which

$$
\frac{1}{f_N \kappa_N^3}=
(\Delta_N - \hbar\omega)^{\frac{3}{2}}\left(\frac{\mid N-\tilde{\sigma}  \mid \pi}{8 e^2 p F^2 d }   \right)^{\frac{1}{2}}
$$
that the stronger the electric field $F$ or/and ribbon width
$d$ is, the larger are the exciton electroabsorption
$c_N^2$ (\ref{E:excfrk}), F-K absorption $c_{N \tiny{(F-K)}}^2$ (\ref{E:frkeld})
and the smaller is the exciton attraction influence $T_N$ (\ref{E:exceff}).
Note that along with this the exciton peak maxima decrease.
The exciton effect increases for frequencies close the exciton peaks
for which $Y_N \ll 1$ and then decreases for intermediate frequencies corresponding to the regions
$0< \kappa_N < \beta_{N0}, \, n+\beta_{Nn} < \kappa_N < n+1,\sim n=0,1,2,\ldots .$
However, even for these frequencies the exciton
effect contributes significantly to the optical absorption.
As the frequency shifts away from the ground exciton peak
$N=n=0$ inside the gap $(\kappa_N \rightarrow 0)$
the exciton effect becomes negligibly small,
$\Gamma^2(-\kappa_N) \simeq 4Y_N^2~\mbox{and}~
c_N^2 \simeq c^2_{N\mbox{\tiny (F-K)}}$.
The spectrum of the exciton and F-K electroabsorption
in the vicinity of the ground exciton peak
and below and above in the intermediate frequency regions
is given in Fig. 4. In this Figure one can see that in accordance
with eqs. (\ref{E:excfrk}) and (\ref{E:exceff}) in the vicinity
of the ground exciton peak the exciton electroabsorbtion
contributes considerably, while the F-K absorption
(\ref{E:frkeld}) is nearly invisible. As mentioned above the chosen
electric field corresponds to a relatively narrow ground peak
and practically unites the first excited broad peak with the region of the
continuous spectrum $\hbar \omega > \Delta_N.$

\begin{figure}[t]
    \centering
    \includegraphics[width=0.75\textwidth]{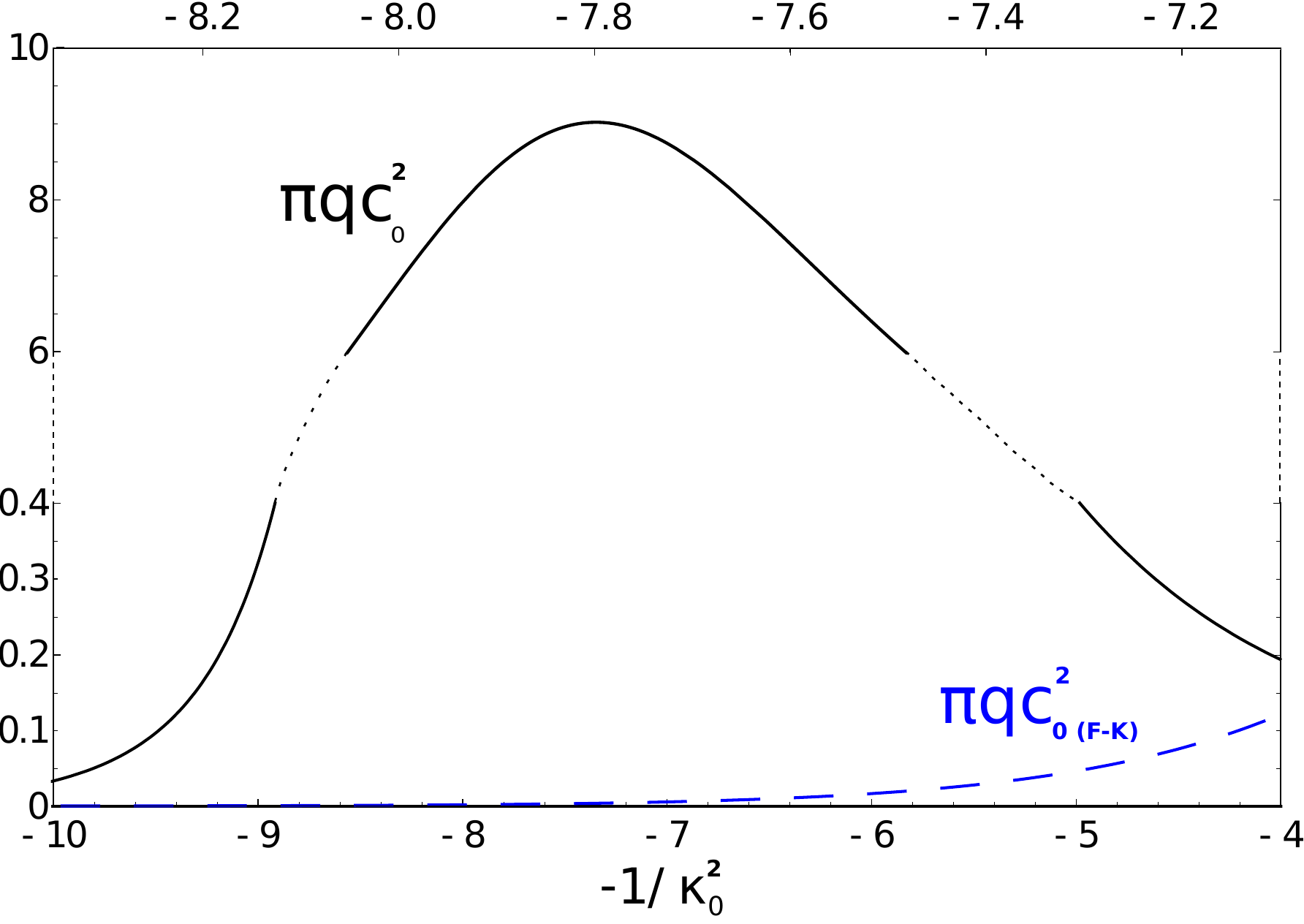}
    \caption{The spectra of the exciton electroabsorption in the vicinity of the ground exciton peak (black line) and of F-K absorption (blue line) as a function of the reciprocal frequency shift
$-\kappa_0^{-2} = Ry_0/(\hbar \omega-\Delta_0 )$ with respect to the
ground threshold $\Delta_0$. The coefficients
$c_0^2~\mbox{and}~c_{0(\mbox{\tiny{F-K}})}^2$, functions
$Y_0~\mbox{and}~\Omega_0$ are calculated from eqs.
(\ref{E:coeff}) and (\ref{E:frkeld}), (\ref{E:Y}) and (\ref{E:exp}), respectively, for
$N=0$. The dimensionless electric field is chosen to be $f_0 = 5$;
sapphire substrate $q=0.24$ is implied.}
\end{figure}

\vspace{1cm}

\noindent \textbf{5.2 Franz-Keldysh exciton absorption} $\hbar\omega > \Delta_N$

\vspace{1cm}

In the frequency region above the edge
of the electroabsorption spectrum is described
by eqs. (\ref{E:excfrkcont}).
The frequency oscillations modulated by the reciprocal square-root factor (1D F-K effect \cite{Ciob}) are
modified by the 1D exciton Sommerfeld factor $Z$
and the exciton phase shift $\Delta\Theta_{NZ}$.
In contrast to semiconductor structures, in particular the bulk
crystals subject to strong magnetic field $B$ \cite{Ciob} and QWR
\cite{Monschm09} in which the oscillation period
depends only on the electric fields $F$,
in the AGNR this period is affected by both the electric
fields $F$ and ribbon width $d$. With increasing each
of these parameters the oscillation period increases, while the phase shift
$\Delta\Theta_{NZ}$ decreases. The Sommerfeld factor demonstrates
that the exciton attraction suppresses the low energy oscillations
and has only a small effect on those of high frequencies positioned away from the edge $\Delta_N$.

It is reasonable to compare the
exciton absorption
dependencies on the electric field $F$ and ribbon width
$d$
with those of the
quasi-1D semiconductor structures i.e.                                                                                                                                                                                                                                                                                                                                                                                                                                                                                                                                                                                                                                                                                                                                                                                                                                                                                                                                                                                                                                                                                                                                                                                                                                                                                                                                                                                                                                                                                                                                                                                                                                                                                                                                                                                                                                                                                                                                                                                                                                                                                                                                                                                                                                                                                                                                                                                                                                                                                                                                                                                                                                                                                                                                                                                                                                                                                                                                                                                                                                                                                                                                                                                                                                                                                                                                                                                                                                                                                                                                                                                                                                                                                                                                                                                                                                                                                                                                                                                                                                                                                                                                                                                                                                                                                                                                                                                                                                                                                                                                                                                                                                                                                                                                                                                                                                                                                                                                                                                                                                                                                                                                                                                                                                                                                                                                                                                                                                                                                                                                                                                                                                                  with the DE \cite{Mon76} and QWR \cite{Monschm09}.
The confinement radius $r_0$ analogical to the
ribbon width $d$ is
the magnetic length
$a_B=\left( \frac{\hbar}{eB}\right)~\mbox{or wire radius}~R$,
respectively. This radius was assumed to be much smaller
than the semiconductor exciton Bohr radius
$a=\frac{4\pi\varepsilon_0 \varepsilon \hbar^2}{\mu e^2}$.
The dependencies of the absorption spectrum on the electric
fields $F$ are common for the AGNR
and semiconductor structures. The weak electric field
does not affect the peak positions
and exciton binding energies, while it
modifies considerably the \emph{shape} of the spectrum. With increasing electric field the exciton peaks become broader
and lower in intensity (eq. (\ref{E:grexc}) ),
absorption grow in the interpeak regions
(eq. (\ref{E:excfrk})) and
frequency oscillations above the edge (eq. (\ref{E:excfrkcont}))
become rare. In contrast to the electric field
effect the influence of the ribbon width $d$ is more pronounced
compared to that of the wire radius $R$ or magnetic field
$B$.

The exciton binding energies  $E_{Nn}^{\mbox{\tiny(b)}}$ in the AGNR
increase linearly with narrowing the ribbon
$E_{Nn}^{\mbox{\tiny(b)}}\sim \Delta_N \sim d^{-1}$,
while in the semiconductor structures this dependence
has the less pronounced logarithmic character
$E_{N0}^{\mbox{\tiny(b)}} \sim \ln^2\frac{r_0}{a}$
for the ground state $n=0$ and is practically
invisible for the excited states $n=1,2,\ldots$.
In the AGNR the confinement dependence of the exciton
peak widths (eq. (\ref{E:grexc}) ) are mostly governed by the exponential factor
$\Gamma_{Nn}\sim d^{-1}
 \exp \left( -\frac{\mbox{const.}}{Fd^2}  \right)$.
This results in a peak widths decrease with the ribbon narrowing.
In the semiconductor structures
the ground peak $n=0$ becomes
 wider with decreasing radius
$r_0~\Gamma_{N0} \sim \ln^2\frac{r_0}{a}$,
while the excited peak widths $\Gamma_{Nn},~n=1,2,\ldots$ remain
constant.  In the AGNR the absorption maxima (\ref{E:grexc})
$\alpha_{Nn}^{\mbox{\tiny(max)}}
\sim \exp \left( \frac{\mbox{const.}}{Fd^2}  \right)$.
In semiconductor structures the confinement
provides only the power growth modified by the logarithmic factor
$\alpha_{Nn}^{\mbox{\tiny(max)}}
\sim \left(r_0 \ln \frac{r_0}{a}  \right)^{-2},\, n\neq 0,
\alpha_{N0}^{\mbox{\tiny(max)}}\sim r_0^{-2}\mid \ln \frac{r_0}{a} \mid^{-1}$. However, there is an exception from the above regarding
the similar dependence of the peak positions.
With increasing the confinement (decreasing the radius $r_0$
and width $d$) the exciton peaks shift towards higher
frequencies in all structures.

In the frequency region below the edge $\hbar \omega < \Delta_N$
the F-K optical absorption in the AGNR (\ref{E:frkeld}) and the effect of the exciton attraction (\ref{E:exceff})
depend on the parameters
$F~\mbox{and} ~ d$ exponentially and by power law, respectively. In the semiconductor structures the coefficient of the
F-K absorption demonstrates the less pronounced power dependence
$\alpha_{\mbox{(\tiny F-K)}}\sim r_0^{-2}$, while the effect of the electron-hole
attraction does not depend on the confinement. In the spectral region
above the edge $\hbar \omega > \Delta_N$ the
oscillations of the F-K absorption (\ref{E:excfrkcont}) {become large in period with widening the AGNR,
while in the semiconductor structures the confinement
does not modify the oscillation period.
\\
\\
\textbf{5.3 The double subband approximation}
\\
\\
Note that all exciton states related to the excited Rydberg series $N\geq 1$ are initially meta-stable. This stems from their coupling to the states of the continuous
spectra emanating from the energetically low-lying thresholds $\Delta_{N-1}$.
At $F=0$ the latter causes the autoionization and transforms the $N\neq 0$ series
of the strictly discrete states into the quasi-discrete
ones (Fano resonances) with the energy widths
$\Gamma_{Nn}^{(\mbox{\tiny{F}})} \sim \Delta_N\sim d^{-1}$
 (eq.(\ref{E:fano})), which increase with decreasing the
 ribbon width $d$ \cite{Monschm12}. The optical absorption coefficient caused by the transitions to these exciton resonant
 states has been calculated in Ref. \cite{Monschm16}. In the
 presence of an electric field $F$ the electroionization
 superimposes on the autoionization that in turn leads to the summation
 of the width $\Gamma_{Nn}^{(\mbox{\tiny {el}})}~\mbox{and}~\Gamma_{Nn}^{(\mbox{\tiny{F}})}$
 determined by eqs. (\ref{E:grexc})
 and (\ref{E:fano}), respectively. However,
 the contribution of each of those widths to the total width
 $\Gamma_{Nn}^{(\mbox{t})}= \Gamma_{Nn}^{(\mbox{\tiny {el}})}
 + \Gamma_{Nn}^{(\mbox{\tiny {F}})}$ is different for the wide and narrow ribbons. The reason for this
is that the widths $\Gamma_{Nn}^{(\mbox{\tiny {el}})}~\mbox{and}~\Gamma_{Nn}^{(\mbox{\tiny {F}})}$ change with varying ribbon
width $d$ in the opposite way. The larger $d$ the less is the
Fano width $\Gamma_{Nn}^{(\mbox{\tiny {F}})}$ and  the width
$\Gamma_{Nn}^{(\mbox{\tiny {el}})}$ becomes larger. Thus, for the narrow AGNR the Fano coupling surpasses the electric field tunneling and
$\Gamma_{Nn}^{(\mbox{\tiny{F}})} > \Gamma_{Nn}^{(\mbox{\tiny {el}})}$.
With further ribbon widening
both effects become equal and then the electric field
ionization dominates that of autoionization. For stronger
electric field $F$ the critical
ribbon width $d_F$ decreases providing the peak widths equality
$\Gamma_{Nn}^{(\mbox{\tiny {F}})} = \Gamma_{Nn}^{(\mbox{\tiny {el}})}$.
As it follows from eqs. (\ref{E:grexc}) and (\ref{E:fano})
the latter condition implies the relationship
$d^2F \simeq \mbox{const.} $ The exciton peak widths
$\Gamma_{Nn}^{(\mbox{\tiny {F}})} ~\mbox{and}~
\Gamma_{Nn}^{(\mbox{\tiny {el}})}$ as a function of the ribbon width
$d$ for the different electric field strengths $F$ are
presented in Fig.5. Note that our results for
the ground $n=0$ exciton state related to the ground
Rydberg series $N=0$ are qualitatively valid for any states
and any series.

\begin{figure}
    \centering
    \includegraphics[width=0.75\textwidth]{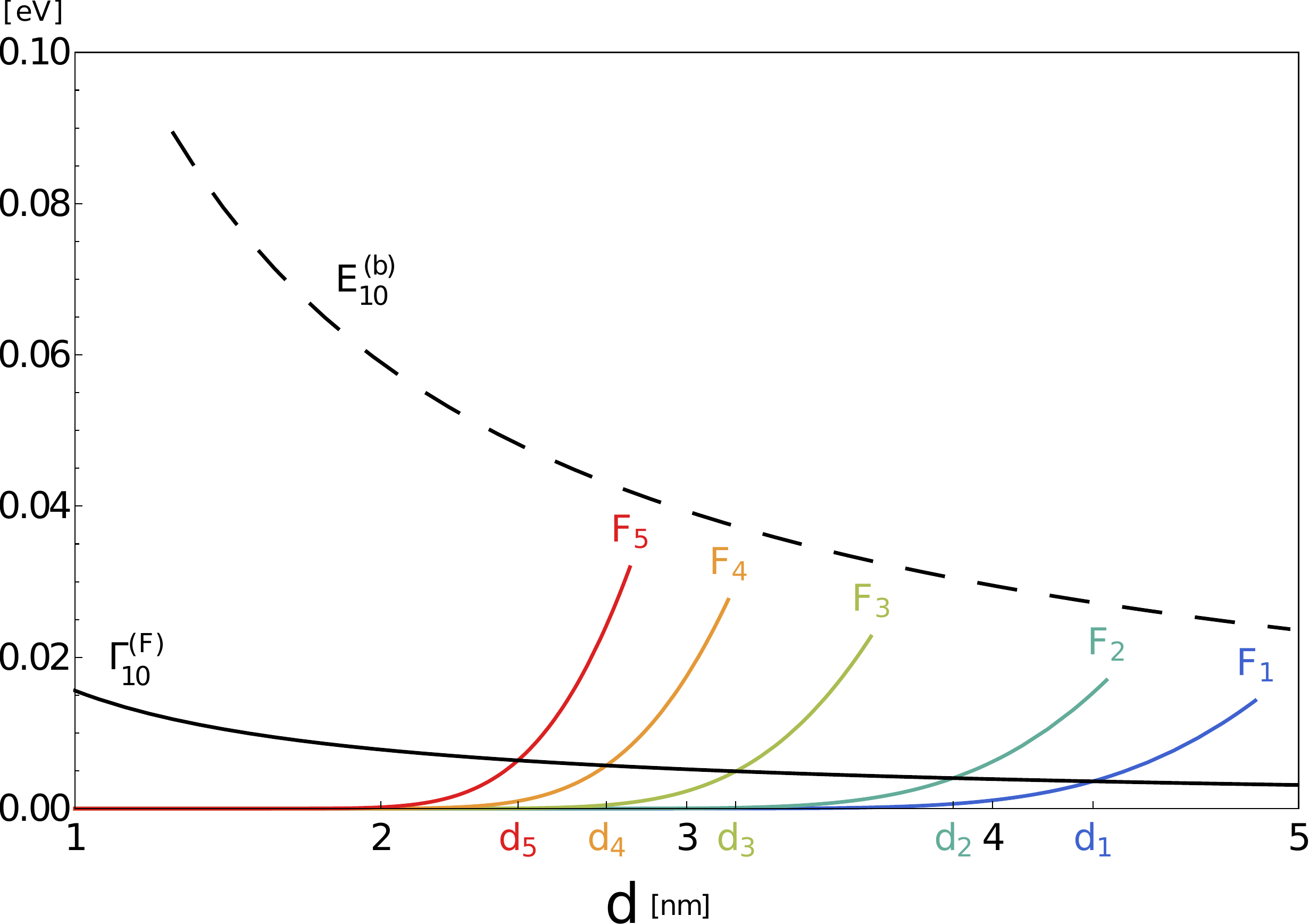}
    \caption{The ground $n=0 $ exciton peak widths
$\Gamma_{10}^{(\mbox{\tiny{el}})}$ (\ref{E:grexc})
corresponding to different electric fields $F_1 =8, F_2 =10, F_3 =15, F_4 =20, F_5=25$ in units kV/cm (color lines),
$\Gamma_{10}^{(\mbox{\tiny{F}})}$ (\ref{E:fano})
induced by intersubband coupling (black line) and the
binding energy $E_{10}^{\mbox{\tiny(b)}}=\mid W_{10} \mid$ (\ref{E:lor}) (dashed line) versus the ribbon width $d$. The first excited Rydberg series $N=1$ is addressed.
The ribbon is placed on the sapphire substrate $q=0.24$}.
\end{figure}

In an effort to connect our results closer to  experiment,
we estimate the expected values for the AGNR with width
4 nm of the family specified by $\tilde{\sigma}=\frac{1}{3}$
and placed on the sapphire substrate $q=0.24$.
The peak positions
$\hbar \omega =E_{Nn}$, binding energies $E_{Nn}^{\mbox{\tiny(b)}}$,
widths $\Gamma_{Nn}^{(\mbox{\tiny {el}})}$, peak absorption maxima
$c_{Nn}^{2\mbox{\tiny (max)}}$, exciton $c_N^2$ and F-K
electroabsorption $c_{N \mbox{\tiny (F-K)}}^2$
coefficients are presented
for the ground $n=0$ and first excited
$n=1$ peaks related to the ground subband $N=0$.
Using
the quantum numbers $\kappa_{0n}$ from eq.
(\ref{E:discen}) we obtain
$E_{00}=316.1~\mbox{meV}$, $E_{01}=338.7~\mbox{meV}$
and $E_{00}^{\tiny \mbox{(b)}}=23.9~\mbox{meV}$,
$E_{01}^{\mbox{\tiny(b)}}=1.38~\mbox{meV}$ for the peak positions
and binding energies, respectively.
Since the ground state binding energy significantly exceeds
that of the first excited state the critical electric field
$\bar{F}_{00}\simeq 24~\frac{\mbox{\tiny{kV}}}{\mbox{\tiny{cm}}}$ is much larger
than $\bar{F}_{01}\simeq 0.31~\frac{\mbox{\tiny{kV}}}{\mbox{\tiny{cm}}}$. These fields
satisfying the condition $\bar{F}_{Nn} = Q_{Nn}d^{-2}$
(eq. (\ref{E:stab})) are the upper bounds of the fields,
providing the complete depletion of the bound exciton
state and the disappearance of the peak from the optical spectrum.
Note that the critical fields providing the equality
of the binding energies $E_{Nn}^{\mbox{\tiny(b)}}$
and spectral widths $\Gamma_{Nn}^{(\mbox{\tiny{el}})}$
can be smaller by a significant factor than those presented above.
In the presence of an electric field
$F= 7~\frac{\mbox{\tiny{kV}}}{\mbox{\tiny{cm}}}$ the ground peak width
becomes $\Gamma_{Nn}^{(\mbox{\tiny{el}})}= 7.56~\mbox{meV}$
yielding for the lifetime $\tau = \hbar/\Gamma_{00}
= 0.082~\mbox{ps}$.
This allows us to conclude that for
weak electric fields of the order of
$F \leq 10~\frac{\mbox{\tiny{kV}}}{\mbox{\tiny{cm}}}$ the
ground exciton
peaks $N\scriptsize{\text{0}}$
remain relatively narrow and
accessible to an experimental study, while for strong fields
of several tens of kV/cm provide a complete
ionization of the exciton states and liberation of the carriers
in the AGNR with widths of several nanometers.

In the spectral region below the edge $\hbar \omega < \Delta_0$
the electron-hole attraction considerably enhances
the absorption for the frequencies positioned far away
from the exciton peaks. It follows from eq. (\ref{E:exceff})
that in the presence of an electric field
$F=7~\frac{\mbox{\tiny{kV}}}{\mbox{\tiny{cm}}}$ the exciton electroabsorption
deeply penetrating into
the gap $(\Delta_0 -\hbar \omega
 \simeq 3.2 E_{00}^{\mbox{\tiny(b)}}\simeq 30 Ry_{0})$
exceeds that of the F-K
by a factor of $T_0 \simeq 8.6$.
In contrast to the discrete part of the spectrum, for frequencies positioned above the edge $\hbar \omega > \Delta_0$
the exciton effect significantly reduces the
magnitude of the
F-K oscillations
which is reflected in the Sommerfeld factor $Z_0$ in eq. (\ref{E:excfrkcont}).
It follows from this equation that
in the frequency region $\hbar\omega - \Delta_0 \simeq 4Ry_0$
and for the electric field $F=7~\frac{\mbox{\tiny{kV}}}{\mbox{\tiny{cm}}}$
the reduction of the oscillating magnitudes is given
by $Z_0 = 0.30$. The corresponding phase is
$\Delta \Theta_{0Z}=-0.44\pi$.

Intending to compare our analytical results with those
calculated numerically we refer to Ref. \cite{gar} in which
the analogous quasi-1D structure, namely the QWR with radius
$R$ subject to an electric field polarized along the wire axis has been studied. The
theoretical approach relied upon the Dirac equation relevant to
the electron (hole) dispersion law
$E(k) = \pm \hbar v_F \sqrt{(2R)^{-2} +k^2}$ ($k$ is the longitudinal
electron (hole) momentum) similar to that for an AGNR \cite{BreyFert}.
The following results for the exciton electroabsorbtion have been
reported. Narrow wires $(R)$ show larger exciton binding energies
$E^{(\tiny{b})}$. Electric fields $(F)$ enhance the subgap $(\hbar\omega < \Delta)$ optical absorption. The electric fields $(F)$ induce the F-K
oscillations. Influence of a given electric field $(F)$ on the excitonic
absorption peak decreases as the Coulomb interactions $(q)$ are made
stronger. Clearly, these conclusions are qualitatively in complete line
with ours (see eqs. (\ref{E:lor}) for $E_{Nn}^{(\tiny{b})} =
- W_{Nn}$, (\ref{E:frkeld}) and Fig.4 for $c_{N \tiny{(F-K)}}^2$,
(\ref{E:excfrkcont}) for F-K oscillations and (\ref{E:exp}) for
$\Omega_N$ and $f_N = F/F_N$ with $F_N \sim q^3$, respectively.) The differences in AGNR and QWR topologies in our paper and in Ref. \cite{gar}, respectively prevents a detailed
quantitative comparison.

As well known, the $\mbox{SiO}_2$ substrate $\varepsilon =3.9$ is widely used in experiments. Nevertheless, we were forced to refrain from the estimates of the possible experimental data related to this material. The problem is here that the corresponding adiabatic parameter
$q=0.37$ is not sufficiently small to provide the needed accuracy implying $q\ll 1$. In principle, the case of similar parameter values
can be investigated by solving the basic set of eqs.
(\ref{E:main}) numerically. Also, we avoid too narrow ribbons with width of 1-2 nm, consisting of several 1D unit cell distances, to be taken
as a candidates for the above-mentioned estimates. In these ribbons its discrete structure manifests itself and the continual approach based on the Dirac equation becomes inappropriate.

The presented analytical results and numerical estimates show that in the AGNR with typical widths and subject to laboratory electric fields the exciton effect manifests itself in the optical absorption
spectrum. An electron-hole attraction on the one hand side significantly modifies the absorption for a very wide frequency region below and above the absorption edge and on the other hand side demonstrates that the exciton peaks are sufficiently narrow and accessible to experimental study. We believe that our developed analytical approach highlights the mechanism of the exciton electroabsorption in AGNR and the auto- and electroionization processes favoring the deliberation of carriers. Also we hope that estimates of the expected experimental values could be usful in further studies
of graphene nanoribbons and their applications in opto- and microelectronics.

\section{Conclusion}\label{S:conc}

In summary, we have developed an analytical
approach to the problem of the exciton
absorption in a narrow armchair graphene
nanoribbon (AGNR) exposed to an external
electric field directed parallel to the graphene
axis. The effect of the strong ribbon
confinement was assumed to be much stronger than
that of the exciton Coulomb electric field, which
in turn considerably exceeds the external field.
In the approximation of the isolated
size-quantized subbands the exciton absorption coefficient
of light polarized parallel to the ribbon axis
has been determined in an explicit form.
We traced the dependencies of the exciton peak positions,
their widths caused by the electric fields and
intensities of the continuous absorption bands
on the ribbon width and electric field strength.
With increasing electric field and widening
the ribbon the exciton peak increases in
 width and decreases in intensity maximum. The exciton
 electron-hole attraction strongly modifies the
 Franz-Keldysh (F-K) absorption related to the
 unbound carriers significantly increasing
 and reducing the F-K absorption in the interpeak
 regions below and above the edge, respectively.
 In the double subband approximation the double channel
 ionization of the exciton discrete states adjacent
 to the first excited size-quantized energy level
 has been considered. These channels are open
 due to the tunneling caused by the electric field
 and autoionization
 induced by the Fano coupling of these states
to the states of the continuous spectrum of the ground subband.
The total exciton peak widths caused by the electro- and autoionization have been determined. The presented analytical results correlate
well with those previously calculated numerically.

Estimates of the expected experimental
values for realistic AGNR and electric fields
yield two aspects of the electric field effect.
Weak electric fields provide a significantly
long exciton lifetime and excitons remain
available for an optical study and use
in optoelectronics. A relatively strong field
ionizing the excitons make the carriers unbound
and promote the AGNR transport properties.

\section{Acknowledgments}\label{S:Ackn}
The authors are grateful to A. Devdariany for useful discussion
and M. Pyzh for significant technical assistance.

\end{document}